\documentclass[%twocolumn,           % Format : 
onecolumn, %twocolumn
               showpacs,            % Pacs : showpacs, noshowpacs
               nopreprintnumbers,     % Preprint: preprintnumbers,
               aps,                 % Society: ...
               prd,          	    % Journal Style : pra, prb, prc, prd, pre,
               letterpaper,             % Size : a4paper, ...
              groupeaddress,      % Affiliation (Title) : groupedaddress,
               nofootinbib,         % Footnote: footinbib, nofootinbib
               tightenlines,        % Remove additional spaces in a line
               showkeys, 11pt,
               notitlepage
               ]{revtex4-1}

\usepackage{graphicx}% Include figure files
\usepackage{dcolumn}% Align table columns on decimal point
\usepackage{bm}% bold math
\usepackage{amsmath}
\usepackage{amsfonts,amssymb}
\usepackage{soul}
\usepackage{color}
\definecolor{v}{rgb}{0.6, 0.2, 0.8} %comentarios VM
\usepackage{xcolor}
\usepackage{enumerate}
\usepackage{ulem}

\newtheorem{de}{Proof}[section] 
\newtheorem{lm}{Lemma}[section]

\begin{document}

\title{\Large{Compact objects in unimodular gravity}}

\author{{J.~A.~Astorga-Moreno}$^1$}
\email{jesus.astorga@alumnos.udg.mx}

\author{{Javier Chagoya}$^{2,3}$}
\email{javier.chagoya@fisica.uaz.edu.mx}

\author{{J.~C.~Flores-Urbina}$^4$}
\email{jfloresu@ipn.mx}

\author{{Miguel~A.~Garc\'ia-Aspeitia}$^{2,3}$}
\email{aspeitia@fisica.uaz.edu.mx}

\affiliation{$^1$Centro Universitario de la Ci\'enega, Universidad de Guadalajara\\
Ave. Universidad 1115, M\'odulo de Investigaci\'on y Tutor\'ias, C.P. 47820 Ocotl\'an, Jalisco, M\'exico.}

\affiliation{$^2$Unidad Acad\'emica de F\'isica, Universidad Aut\'onoma de Zacatecas, Calzada Solidaridad esquina con Paseo a la Bufa S/N C.P. 98060, Zacatecas, M\'exico.}

\affiliation{$^3$Consejo Nacional de Ciencia y Tecnolog\'ia, \\ Av. Insurgentes Sur 1582. Colonia Cr\'edito Constructor, Del. Benito Ju\'arez C.P. 03940, Ciudad de M\'exico, M\'exico.\renewcommand{\andname}{\ignorespaces} }

\affiliation{$^4$Unidad Profesional Interdisciplinaria de Ingenier\'ia Campus Zacatecas. \\ Instituto Polit\'ecnico Nacional.
Blvd. del Bote S/N Cerro del Gato Ejido, La Escondida, C.P.98160 Zacatecas, M\'exico.}

%-------------------------------------------------------------------------------------------------
%-------------------------------------------------------------------------------------------------
\begin{abstract}
Unimodular gravity provides a theoretical framework that allows for
non-conservation of energy-momentum, with possible implications for 
the cosmological constant problem. It is then important to study the 
predictions of unimodular gravity in other gravitational regimes. In this work we study stellar dynamics under the assumption of non-conserved energy-momentum. We find that constant density objects can be as compact as Schwarzschild black holes. For polytropic objects, we find
modifications due to the non-conservation of energy-momentum that 
lead to sizeable effects that could be constrained with observational data. 
Additionally, we revisit and clarify the Reissner-Nordstr\"om solution in unimodular gravity.   
We also study
gravitational collapse and discuss possible implications for the growth of structure.    
\end{abstract}

\keywords{Unimodular gravity, stellar dynamics.}
%\draft
\pacs{}
\date{\today}
\maketitle
%%%%%%%%%%%%%%%%%%%%%%%%%%%%%%%%%%%%
\section{Introduction}
%%%%%%%%%%%%%%%%%%%%%%%%%%%%%%%%%%%%
Stellar dynamics is an outstanding laboratory to explore the predictions of
general relativity (GR) and alternative theories in the strong field regime.
A variety of systems, such as dwarf stars, neutron stars, or even collapsed objects 
like black holes offer observational windows to constrain the theory of gravity.
This has led to the study of astrophysical systems in several alternatives to
GR, for example, $f(R)$ theories \cite{Sotiriou:2008rp,Canate:2015dda}, brane-world models \cite{Germani,*Maartens:2010ar,*Garcia-Aspeitia:2014,*Linares:2015,*Garcia-Aspeitia:2015mwa,*Visinelli:2017bny}, noncommutative theories \cite{Bertolami:2009wa,Garcia-Aspeitia:2015zpf}, scalar and vector-tensor theories \cite{Chagoya:2014,Cisterna:2015yla,Maselli:2016gxk,Chagoya:2016aar,Minamitsuji:2016ydr,Chagoya:2017fyl,Filippini:2017kov,Chagoya:2018lmv,Valdez} etc.
In particular, it is important to study how the equilibrium equations are modified
and what are the consequences for the magnitudes that characterize compact systems, 
such as the mass, radius, and compactness. For simplicity these studies often begin with
constant density objects, but it is also important to use more realistic equations of 
state and identify degeneracies between modifications due to the theory of gravity and 
those due to the equation of state. 
On the other hand, the long-standing cosmological constant problem \cite{Weinberg,Zeldovich} continues to be
a motivation for proposing alternative theories of gravity. One of these alternatives, known as Unimodular Gravity (UG) \cite{Einstein:1919gv,Ng:1990rw,doi:10.1063/1.529283}, attempts to relax this problem by a mechanism that
makes vacuum energy non-gravitating and attributes the observed cosmological constant
to an arbitrary integration constant of the theory, helping to evade the gap between the theoretical and observational estimates of the cosmological constant. This form of UG
works under the assumption that the energy-momentum tensor is covariantly conserved, and
does not offer physical insights on the nature of the cosmological constant: its classical field equations are the same as in GR  \cite{Smolin:2009ti,Ellis, Gao:2014nia,Basak:2015swx,Padilla:2014yea}. The study of cosmological perturbations 
requires some care in the gauge choice, for instance, now the gauge symmetry consists of transverse diffeomorphisms only, but it has been shown that in gauge invariant
quantities the equivalence between GR and UG holds \cite{Basak:2015swx}. 
At the quantum level, this equivalence is not yet clear. 
Differences have been
argued to arise, with implications for the hierarchy and 
radiative stability of the cosmological constant \cite{Smolin:2009ti,Eichhorn:2013xr,Alvarez:2015sba,Ardon:2017atk}, but it has also been 
suggested that the equivalence depends on the details of the classical starting point -- whether the determinant of the metric is explicitly constrained or not --
as well as on the quantization procedure \cite{Bufalo:2015wda,Percacci:2017fsy}.

As mentioned above, the conservation of the energy-momentum tensor is not automatic 
in UG but is introduced as an additional assumption. Recently, the possibility to 
discard 
this assumption has received some attention in the literature, partially motivated
by novel ideas proposed by Perez et al. \cite{Perez:2017krv,Perez:2018wlo} that 
do offer an insight on the nature of the cosmological constant by allowing for a
non-conserved energy-momentum tensor, a feature that can be incorporated in UG.
Further, exhaustive studies on the cosmological implications of this version of UG 
have been
reported in \cite{Daouda:2018kuo} and \cite{ Garcia-Aspeitia:2019yni}. 

Given this scenario, it is important to put UG to test also in extreme gravitational
laboratories, such as stellar dynamics, with the assumption that energy-momentum
is non-conserved, thus stopping the classical theory from automatically reducing to GR.
In this work we study static, spherically symmetric solutions to the UG field equations
subject to an additional condition that is required in order to close the system of equations, 
since in UG the number of independent field equations is reduced by one due to the trace-free 
property of the field equations. We exploit this additional condition in two ways, first to
simplify the system of equations and obtain some analytical results, and then to parameterize
the type of non-conservation of energy-momentum in our numerical results and contrast with GR
predictions.

This paper is organized as follows: In Sec.~\ref{sec:UG}, we briefly review the
theoretical framework of UG. Sec.~\ref{sec:sssug} is dedicated to the study of static, spherically symmetric solutions. We begin by exploring the relation between solutions for metrics that satisfy explicitly the unimodular condition -- constant metric determinant -- and solutions
that do not satisfy this condition, we justify that the systems that we explore in this work
can be analyzed in either coordinate chart. We then study an analytic solution of UG through the imposition of a simplifying ansatz that allows us to obtain the Tolman-Oppenheimer-Volkoff (TOV) equation, which we use to analyze some properties of a star in this conditions. In Sec.~\ref{sec:tovconstant} we study constant density configurations under
an assumption for the non-conservation of energy-momentum; we perform a numerical analysis and
explore the behaviour of our solutions near the Buchdahl limit of GR. In Sec.~\ref{sec:poly}
we extend the previous numerical analysis to account for stars described by a polytropic equation of state (EoS), obtaining modifications that could provide constraints on the non-conservation of energy-momentum. Finally, in Sec.~\ref{GC} we study gravitational collapse, showing that collapse times get modified, and we  discuss some consequences for black hole and structure formation in our Universe. Sec.~\ref{Sec:CO} is
devoted to conclusions and perspectives for the study of stellar dynamics in UG.
In addition, we include several
appendices with details of our calculations and assumptions:  we discuss the equivalence between unimodular and FLRW-like metrics; we give a formal justification for the simplifying assumptions used in Sec.~\ref{sec:sssug}, we provide details of the equations of motion for constant density objects, we
discuss how different assumptions for the non-conservation of energy-momentum affect our results, and we revisit the Reissner-Nordstr\"om solution, clarifying misleading 
results reported in previous studies. 

%%%%%%%%%%%%%%%%%%%%%%%%%%%%%%%%%%%%
\section{Unimodular gravity: Lagrangian and equations of motion} \label{sec:UG}
%%%%%%%%%%%%%%%%%%%%%%%%%%%%%%%%%%%%
Unimodular gravity can be described by the  action 
\begin{equation}
S = \int d^4x \sqrt{-g} \left(\frac{1}{2\kappa} R + \mathcal L_{matter}  \right)\, ,\label{eq:eh}
\end{equation}
where $\kappa=8\pi G$ with $G$ the Newtonian gravitational constant. Importantly, the metric 
determinant in~\eqref{eq:eh} is restricted to satisfy the {\it{unimodular condition}} $\sqrt{-g} = \epsilon_0$, where $\epsilon_0$
is a fixed scalar density usually set to unity, i.e $\sqrt{-g}=1$. The unimodular action can then be written as
\begin{equation}
S = \int d^4x \epsilon_0 \left(\frac{1}{2\kappa} R + \mathcal L_{matter}  \right)\, , \label{eq:ugaction}
\end{equation}
where all the tensors in the action are constructed
with a metric that satisfies the unimodular condition. After some manipulations --  transparent when
the unimodular condition is incorporated into the action by means of a Lagrange multiplier -- the equations of motion
result in
\begin{equation}
\xi_{\mu\nu}:= G_{\mu\nu} - \kappa T_{\mu\nu} =- \frac{1}{4} g_{\mu\nu} ( R +   \kappa  T  )\, ,
\label{eq:ugeom}
\end{equation}
where $T_{\mu\nu}$ is the standard energy-momentum tensor
\begin{equation} T_{\mu\nu} = -\frac{2}{\sqrt{-g}} \frac{\delta(\sqrt{-g} \mathcal{L}_{matter})}{\delta g^{\mu\nu}}\, . \label{eq:tmunu}
\end{equation}
Some confusion might arise here since, strictly, the 
energy-momentum tensor in UG should be defined like in~\eqref{eq:tmunu} but without
the metric determinants. However, it can be shown that the
combination $T_{\mu\nu} - g_{\mu\nu}T/4$ which appears in
the equations of motion is independent of what definition of the energy-momentum tensor is used~\cite{Smolin:2009ti}.
In contrast to General Relativity, where the trace of the equations of motion
gives $R = - \kappa T$, here the equations of motion are trace-free, and the differences with respect
to GR are indeed parameterized by $R + \kappa T$. Furthermore, while in GR
the Bianchi identities $\nabla^\mu G_{\mu\nu} = 0$ enforce
the covariant conservation of the energy-momentum tensor, i.e. $\nabla^\mu T_{\mu\nu} =0$, in unimodular
gravity there is the possibility to have non-conserved energy-momentum tensors since the conservation that must be 
satisfied is
\begin{equation}
\nabla^\mu \left(\kappa T_{\mu\nu} - \frac{\kappa}{4} g_{\mu\nu} T - \frac{1}{4} g_{\mu\nu} R  \right) = 0\, . \label{eq:enmom}
\end{equation}
If $\nabla^\mu T_{\mu\nu} = 0 $ is assumed, then the above equation implies $\partial_\nu ( \kappa T + R) = 0$,
so that we can write $ \kappa T + R = -4\Lambda$ for some integration constant $\Lambda$. Plugging this
back in the equations of motion of unimodular gravity we get 
$G_{\mu\nu} +\Lambda  g_{\mu\nu} = \kappa T_{\mu\nu}$. Then, assuming conservation of the energy-momentum
tensor, the equations of motion of GR with a cosmological constant are recovered. 

In the formulation of UG described above the unimodular condition is reflected in 
the trace-free property of the equations of motion. For our purposes this formulation is enough, since what we actually exploit in this work is the fact that this condition reduces the number of 
independent equations of motion,
but it is worth mentioning that there
are other approaches in the literature, for instance, the field equations of UG can be 
derived from a fully diffeomorphism invariant action~\cite{Henneaux:1989zc}. 

Rather than assuming~$\nabla_\mu T^{\mu}{}_\nu =0$, which automatically
leads to the usual Einstein field equations, in the next
sections we work
with~\eqref{eq:ugeom} and look for configurations where
the energy-momentum tensor is not conserved.
Physical motivations for this possibility have been presented in~\cite{Josset:2016}.

%%%%%%%%%%%%%%%%%%%%%%%%%%%%%%%%%%%%
\section{Static, spherically symmetric solutions} \label{sec:sssug}
%%%%%%%%%%%%%%%%%%%%%%%%%%%%%%%%%%%%
Before we start, it is important to mention that the unimodular condition (hereafter, we consider the unimodular condition as $\sqrt{-g} = 1$ where $\epsilon_0=1$) is not the most formal way to define unimodular gravity: what is really relevant
is that the equations of motion are obtained by considering an invariant volume form. A volume form is coordinate independent,
while $\sqrt{-g} = 1$ is not. The  physical consequence of the restricted variation considered in unimodular gravity
is the fact that the equations of motion are trace-free. At the level of the equations of motion we can impose any ansatz for the metric; furthermore, at least locally, any metric can be rewritten in a form that satisfies $\sqrt{-g} = 1$.

Nevertheless, for the sake of completeness and clarity, here we review some static, spherically symmetric solutions
both in unimodular coordinates (i.e., coordinates where the unimodular condition is satisfied explicitly) and in standard
spherically symmetric coordinates.  In addition, in Appendix \ref{app:eq} we demonstrate the equivalence between a FLRW metric\footnote{We use for simplicity the FLRW line element in order to illustrate the equivalence between metrics, due to the integrability in this particular case.} in its standard form and in a form that fulfills the unimodular condition, showing that it is irrelevant which metric we are using and the physical results
are the same in each system. Another important point that it is necessary to remark is that UG coordinates does not have advantages from a numerical point of view\footnote{Unless it helps to diagonalize the metric, which is not the case with this paper.} and thus, the following analysis has only the goal to underscore the connection between both coordinates system. Let us begin our discussion in unimodular coordinates.

%%%%%%%%%%%%%%%%%%%%%%%%%%%%%%%%%%%%
\subsection{Static, spherically symmetric solutions of GR in unimodular coordinates} \label{subsec:sssug}
%%%%%%%%%%%%%%%%%%%%%%%%%%%%%%%%%%%%
Given a metric of the form
\begin{equation}
ds^2 = -f(r)dt^2 + h(r)^{-1}dr^2 + r^2(d\theta^2 +\sin^2\theta d\varphi^2)\, ,\label{eq:sssmetric}
\end{equation}
we can perform a coordinate transformation $dr =  \sqrt{h(r)/(r^4f(r))}  dy $, $x = \cos\theta$, such that the metric
rewrites as.
\begin{equation}
ds^2 = -f(y)dt^2 +  \frac{dy^2}{r(y)^4 f(y)}  + \frac{r(y)^2dx^2}{1-x^2} + r(y)^2 (1-x^2) d\varphi^2\, , \label{eq:ucoords}
\end{equation}
as long as the radial coordinate can be expressed in terms of $y$. For example, for a solution 
with $f(r) = h(r)$ the coordinate change simplifies to $r = (3y)^{1/3}$. Using this, the Schwarzschild solution
in unimodular coordinates reads\,(see \cite{Torres:2017ygl} for a more complete study of this solution in UG)
\begin{eqnarray}
ds^2 &&= -\left(1-\frac{2M}{ (3y)^{1/3}}\right)dt^2 + \left(1-\frac{2M}{(3y)^{1/3}}\right)^{-1} \frac{dy^2}{(3y)^{4/3}} \nonumber 
\\ && \hspace{1em}+ \frac{ (3y)^{2/3}dx^2}{1-x^2} + (3y)^{2/3} (1-x^2) d\varphi^2\, .
\end{eqnarray}
A similar procedure can be applied to (anti)-de Sitter and Reissner-Nordstr\"om metrics (the latter is analyzed in Appendix~\ref{app:rn}). When $f\neq h$, expressing $r$ in terms
of $y$ becomes more complicated, for example, for a constant density TOV solution to the Einstein-Hilbert
equations with matter described by  $T^{0}_{0} = -\rho_0, T^i{}_{i} = p(r)$ and all other elements vanishing, the metric can be 
written in the spherical coordinates~\eqref{eq:sssmetric} with
\begin{eqnarray}
f(r) && =   \frac{\rho_0^2}{(\rho_0 + p(r))^2}\, , \nonumber \\
h(r) && =  1 - \frac13 r^2 \kappa \rho_0\, , \nonumber \\
p(r) && =  \rho_0 \frac{\sqrt{R_s^2 \kappa \rho_0 -3} - \sqrt{r^2 \kappa \rho_0 -3}}{  \sqrt{r^2 \kappa \rho_0 -3} - 3\sqrt{R_s^2 \kappa \rho_0 -3}}\, ,
\end{eqnarray}
where the constant $R_s$ is the radius of the compact object, defined by the vanishing of $p(r)$. The change of coordinates
requires us to integrate 
$$
dy =  -\frac{\sqrt{3} r^2 \left(\sqrt{3-r^2 \kappa  \rho _0}-3 \sqrt{3-\kappa  R_s{}^2 \rho _0}\right)}{2 \sqrt{\left(3-r^2 \kappa  \rho _0\right) \left(3-\kappa  R_s{}^2 \rho _0\right)}} dr\, ,
$$
 and then solve for $r$ as a function of $y$. In a small $\rho_0$ approximation (formally defined by introducing a small, dimensionless parameter $\epsilon$ such that
 $\rho_0 \to \epsilon\rho_0$), we have to solve
 \begin{equation}
 \frac{r^3}{3}+\frac{1}{180} r^3 \kappa  \left(9 r^2-5 R_s{}^2\right) \rho _0+\frac{r^3 \kappa ^2 \left(9 r^4-7 R_s{}^4\right) \rho _0^2}{1008}+\frac{5 r^3 \kappa ^3 \left(r^6-R_s{}^6\right) \rho _0^3}{2592}+\mathcal O(\rho^4) = y\, . \label{eq:yofr}
 \end{equation}
 Since $y$ is a new coordinate it does not depend on $\rho_0$, so we 
can take $r = r_0(y) + \rho_0 r_1(y) + \rho_0 r_2(y) + \dots$.  Proceeding in this way we find a perturbative solution to arbitrary order in
 $\rho_0$. The first few terms read
 \begin{eqnarray}
r_0(y) && = (3 y)^{1/3} \, , \nonumber \\
r_1(y) && = \frac{1}{180} \left( 5\ 3^{1/3} y^{1/3} \kappa  R_s{}^2   -27 y \kappa \right)\, , \nonumber  \\
r_2(y) && = \frac{9\ 3^{2/3} y^{5/3} \kappa ^2}{2800}-\frac{1}{40} y \kappa ^2 R_s{}^2+\frac{11 y^{1/3} \kappa ^2 R_s{}^4}{432\ 3^{2/3}}\, .
\end{eqnarray} 
Notice that $r_0(y)$ coincides with the coordinate transformation of a vacuum solution. We can now write down
the unimodular form of the TOV metric for small, constant density objects in GR:
\begin{eqnarray}
f(y) && % \frac{\rho^2}{(\rho_0 + p[r(y)])^2}
 \approx 1+\frac{1}{6} \left(3^{2/3} y^{2/3} \kappa -\kappa  R_s{}^2\right) \rho _0+\frac{\left(27\ 3^{1/3} y^{4/3} \kappa ^2+50\ 3^{2/3} y^{2/3} \kappa ^2 R_s{}^2-75 \kappa ^2 R_s{}^4\right) \rho _0^2}{2160}+\dots  \, \nonumber \\
 p(y) &&\approx\frac{1}{12} \kappa  \left(R_s{}^2-3^{2/3} y^{2/3}\right) \rho _0^2+\frac{\kappa ^2 \left(27\ 3^{1/3} y^{4/3}-35\ 3^{2/3} y^{2/3} R_s{}^2+30 R_s{}^4\right) \rho _0^3}{1080}+\dots\, .
\end{eqnarray}
The radius of the star is given by $y(R_s)$ and can be obtained from~\eqref{eq:yofr}.
 The previous results verify that both Schwarzschild and the TOV metric of a constant density object can be expressed in unimodular coordinates. Given that Schwarzschild is a vacuum solution and that the energy-momentum tensor of TOV  satisfies $\nabla^{\mu} T_{\mu\nu} = 0$, they have to be also solutions of unimodular gravity, although not the most general ones since in UG
with  $\nabla^{\mu} T_{\mu\nu} = 0$ there is an additional integration constant that embeds these solutions in 
(anti-)de Sitter space-time.  Now that we are convinced that a change of coordinates from standard to unimodular coordinates
exists also for solutions in presence of matter, we continue our study of static, spherically symmetric solutions of
unimodular gravity in standard coordinates, allowing for a non-conserved energy-momentum tensor. 

%%%%%%%%%%%%%%%%%%%%%%%%%%%%%%%%%%%%
\subsection{Static, spherically symmetric solutions of UG} \label{subsec:sss}
%%%%%%%%%%%%%%%%%%%%%%%%%%%%%%%%%%%%
In order to study stellar dynamics, we assume that the geometry is
described by the spherically symmetric line element given in Eq.~\eqref{eq:sssmetric}, and we consider a matter sector characterized by a perfect fluid whose energy-momentum tensor is expressed in the form
\begin{equation}
T_{\mu\nu}=\rho u_{\mu}u_{\nu}+p (g_{\mu\nu}+u_{\mu}u_{\nu}),    \label{fluid}
\end{equation}
where $p=p(r)$ and $\rho=\rho(r)$ are, respectively, the pressure and density of the stellar matter of interest, $u_{\mu}$ is the fluid four-velocity, which  satisfies the condition $g_{\mu\nu}u^{\mu}u^{\nu}=-1$, and $g_{\mu\nu}+u_{\mu}u_{\nu}$ is orthogonal to $u_{\mu}$. In the following, 
we derive analytic expressions for the masss and gravitational energy of gravitationally bound objects under some assumptions that allow for analytical progress.

%%%%%%%%%%%%%%%%%%%%%%%%%%%%%%%%%%%
\subsubsection{ Analytic solution with a particular ansatz}\label{sec:an1}
%%%%%%%%%%%%%%%%%%%%%%%%%%%%%

This exercise is aimed to obtain an analytic solution of the UG field equations in order to
gain some insight on the physics that happens in this context, and later on extend this knowledge to the numerical solutions. For this task, we assume 
that either $R_{tt}=0$, $R_{rr}=0$ or $R_{\theta\theta}=0$. This ansatz
simplifies the field equations and allows us to obtain analytic results (see Appendix \ref{Cdem} for a formal justification of this ansatz choice). 

The field equations give us the form of $h(r)$ as
\begin{equation}
    h(r)=1-\frac{2G\mathcal{M}_{UG}(r)}{r}\, ,\label{eq:hmass}
\end{equation}
where we define
\begin{equation}
    \mathcal{M}(r)_{UG}\equiv\mathcal{C}\int_0^r4\pi r^{\prime 2}(p+\rho)dr^{\prime}\, , \label{Mass}
\end{equation}
with $\mathcal C$ a constant that depends on which component of
the Ricci tensor is set to zero: $\mathcal{C}=3/2$ for $R_{tt}=0$ and 
$\mathcal{C}=1/2$ both for $R_{rr}=0$ and $R_{\theta\theta}=0$.
After some manipulations that combine the field equations with eq.~\eqref{eq:enmom}, we get
\begin{equation}
    \frac{f(r)'}{f(r)}=\pm\frac{p(r)'+\rho(r)'}{p(r)+\rho(r)}\, , \label{B}
\end{equation}
and with this we can arrive to an equation that contains only
the mass function, the pressure, and the density of matter, 
this is the modified TOV equation in UG under the assumptions mentioned 
above:
\begin{eqnarray}
        -r^2(p'+\rho')=G  \mathcal{M}_{UG} \rho \left[1+\frac{p}{\rho}\right] \left[\pm\mathcal{S}\frac{4\pi r^3(p+\rho)}{\mathcal{M}_{UG}}\mp2\right] \left[1-\frac{2G\mathcal{M}_{UG}}{r}\right]^{-1}\, , \label{TOV}
\end{eqnarray}
where $\mathcal{S}=1$ for $R_{tt}=0$ and $\mathcal{S}=3$ both for 
$R_{rr}=0$ and $R_{\theta\theta}=0$.
One concern is that the TOV equation found in this approach is not
continuously connected to the 
Newtonian equation found in the weak field limit of GR. This is 
caused by the ansatz imposed to integrate and obtain the analytic solution: 
by choosing these ansatz we are imposing a non-infinitesimal deviation 
from GR. It is interesting that the system admits solutions under these
conditions and this could lead to observable effects useful to constrain the 
model.

\begin{figure}
\centering
\par\smallskip
{\includegraphics[width=0.75\textwidth]{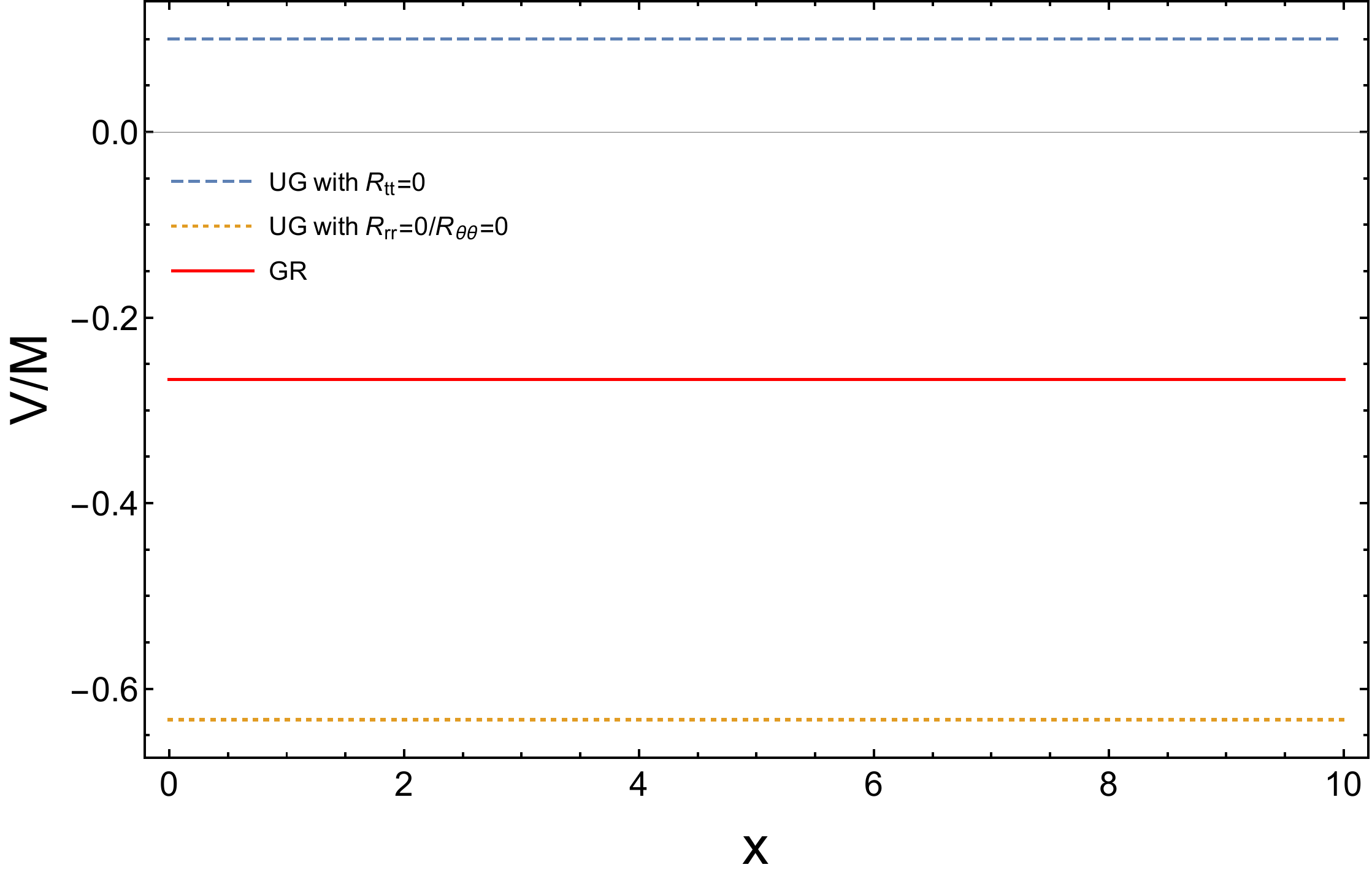}}
\caption{Behavior of gravitational energy described by Eq. \eqref{pot}, for the two ansatz imposed and a comparison with the standard behavior of GR. As a initial conditions, we propose $\bar{\rho}=1$ and $p\ll\rho$.}
\label{Figure:V}
\end{figure}

Integrating Eq.~\eqref{B} and using the modified TOV equation, we  obtain the $g_{tt}$ component of the metric,
\begin{eqnarray}
    f(r)=\exp\left\lbrace\pm\int_r^{\infty}\frac{2G}{r^{\prime 2}}[\mp\mathcal{M}_{UG}\pm2\pi\mathcal{S} r'^{3}(\rho+p)]\left[1-\frac{2G\mathcal{M}_{UG}}{r'}\right]^{-1}\right\rbrace dr^{\prime},
\end{eqnarray}
where we consider the boundary condition $f(\infty)=1$ in order to obtain an asymptotic Minkowski space-time. The upper signs corresponds to $R_{tt}=0$ 
and the lower signs to the other two cases. Outside the configuration
of matter the pressure and density vanish and the Schwarzschild solution 
is recovered, this is possible since our ansatz is automatically satisfied
for Schwarzschild. As a complement, we compute the redshift of spectral lines from the surface of the star as
\begin{equation}
z+1=\left(1-\frac{2M_{UG}G}{R}\right)^{1/2}, \label{eq:surfz}
\end{equation}
where $M_{UG}\equiv\mathcal{M}(R)_{UG}$, given by Eq. \eqref{Mass}, existing a substantial change in comparison with the standard result due to the presence of $p$ and $\mathcal{C}$ in the previous equations. On the other hand, in similarity with GR, we expect that the number of nucleon in the star can be written as \cite{weinberg1972gravitation}
\begin{equation}
N=\int_0^R4\pi r^2\left[1-\frac{2G\mathcal{M}_{UG}(r)}{r}\right]^{-1/2}n(r)dr,
\end{equation}
being $n(r)$ the proper number density. In addition, the internal energy of the star is given by $E\equiv M-m_NN$, where $m_N=1.66\times10^{24}g$ is the rest mass of a nucleon. If we now assume a proper internal material energy density \footnote{The form of $e(r)$ is inspired by the structure of the UG mass given in Eq.~\eqref{Mass}, which depends both on $p$ and $\rho$, in contrast with the mass and $e(r)$ for GR which depend only on $\rho(r)$ (see \cite{weinberg1972gravitation} for details in the GR case).} $e(r)\equiv(\rho(r)+p(r))-m_Nn(r)$, we have $E=T+V$, where
\begin{eqnarray}
    T&=&\int_0^R4\pi r^2\left\lbrace1+\frac{G\mathcal{M}(r)_{UG}}{r}+...\right\rbrace e(r)dr, \\ V&=&-\int_0^R4\pi r^2\left\lbrace(1-\mathcal{C})+\frac{G\mathcal{M}(r)_{UG}}{r}+\frac{2}{3}\left(\frac{G\mathcal{M}(r)_{UG}}{r}\right)^2+...\right\rbrace(\rho(r)+p(r))dr \label{V},
\end{eqnarray}
 where $T$ and $V$ are the thermal and gravitational energies in UG, respectively.
In order to compare Eq. \eqref{V} with the standard gravitational energy, we propose the following dimensionless variables
\begin{equation}
\tilde{V}=V/M, \;\; \bar{\rho}=\rho/\rho_{eff}, \;\; \bar{p}=p/\rho_{eff}, \;\; {\rm and} \;\; x=\sqrt{GM/R}(r/R), 
\end{equation}
with $\rho_{eff}=3M/4\pi R^3$, $M$ the standard GR mass, and assuming the case where $p\ll\rho$ for a constant $\rho$. Therefore we finally have
\begin{equation}
\tilde{V}\approx-\bar{\rho}(1-\mathcal{C})-\frac{3}{5}\bar{\rho}^2C^m_{comp}, \label{pot}
\end{equation}
where $C_{comp}^{m}\equiv GM\mathcal{C}/R$ is the stellar compactness 
modified by UG together with the ansatz made in this section for the Ricci tensor. Eq.~\eqref{pot} is to be compared to the dimensionless gravitational energy in GR, $\tilde{V}=-(3/5)\bar{\rho}^2C_{comp}$, which is shown in Fig.~\ref{Figure:V}  assuming $\bar{\rho}=1$, notice that in this particular case the traditional GR behavior is recovered when $\mathcal{C}=1$, therefore $C_{comp}^m\to C_{comp}=GM/R$. However,
while in GR the constraint $GM/R<4/9$ holds, it remains to be seen whether UG allows for higher
values of the compactness, this is analyzed in the next section. Also, notice that in some cases UG predicts positive values for the gravitational energy, this suggests that the ansatz  $R_{tt}=0$ has no physical interpretations. For the other ansatz we found a lower gravitational energy than the one expected in GR. This result highlights
the differences due to the UG modifications
to the dynamical equation. 

In the next section we present a full
numerical study of compact objects in UG without assuming the ansatz of this section, but rather closing the system of equations with 
an assumption on the type of violation of
energy-momentum tensor that allows for a continuous limit to GR. 

%%%%%%%%%%%%%%%%%%%%%%%%%%%%%%%%
\section{On the choice of non-conservation of $T_{\mu\nu}$}\label{sec:vio}
%%%%%%%%%%%%%%%%%%%%%%%%%%%%%%%%

The trace-free property of the equations of motion in UG reduces the number 
of independent equations of motion. In GR, when a matter Lagrangian and a spherically symmetric ansatz for the metric are considered, the metric equations of motion contain three independent equations -- one of them
equivalent to the Bianchi identities combined with the conservation of $T_{\mu\nu}$. Together with an EoS, these equations suffice to determine
the four free functions of the system (two metric functions and the pressure and density of matter). Under the same considerations, in UG there are only two independent metric equations of motion, which together with the EoS can
determine only three of the four free functions. Therefore we need an extra 
condition to close the system of equations. One possibility is to use this
freedom to impose simplifying assumptions on the equations of motion, as 
in Sec. \ref{sec:an1}.
Another option is to impose a form of violation of $\nabla_\mu T^{\mu}_{\nu} =0$. This has the advantage that we have under control the non-conservation
of $T^\mu{}_\nu$ in the model. Also, we can parametrically recover the
GR solutions when this non-conservation is small. But, what type
of non-conservation should we choose? Most studies on UG 
actually impose conservation of $T_{\mu\nu}$ and take the ``automatic''
presence of a cosmological constant as an integration constant as the main
characteristic of UG. However, this approach leads to the same dynamics as GR
with cosmological constant~\cite{Padilla:2014yea}. On the other hand, a few mechanisms leading to non-conservation
of  energy momentum have been discussed in the literature. For instance,
in \cite{Perez:2017krv} non-conservation originating from the interaction
between a discrete space-time and matter at a microscopic level is 
seen as a friction-like force acting on massive particles, which
macroscopically generates an accelerated expansion of the universe. Similar energy-momentum diffusion effects have been found in causal set theory
\cite{Dowker:2003hb,Philpott:2008vd}.  Another scenario
where non-conserved energy-momentum arises is found in non-unitary
modifications of quantum dynamics, such as the {\it{Continuous Spontaneous
Localization}} (CSL) collapse model. Recently, NSs have been pointed out as
competitive candidates to test this model~\cite{Tilloy:2019nnp}. Dissipative
effects can also be motivated by standard physics of NSs. After they form, NSs undergo a rapid cooling phase driven by the Urca process, in which neutrinos escape from the star carrying energy away\cite{Lattimer:1991ib,Brown:2017gxd}. After a few minutes,
the Urca process is replaced by a much slower modified version that lasts 
up to a million years (see, e.g., the {\it{Minimal cooling paradigm}}~\cite{Page:2004fy}). Along the same lines, with exception of the CSL model,
all the interpretations above lead to energy-momentum loss, and would 
therefore correspond to a negative  $k$ in our ansatz for non-conservation.  A detailed incorporation of these effects in modified gravity is an interesting line of research that seems natural to pursue in UG, since the framework allows for a non-conserved $T_{\mu\nu}$, which is incompatible with GR.
 Given the limited number of works in this direction (see \cite{Daouda:2018kuo,Garcia-Aspeitia:2019yni}
for a cosmological study), it seems better to start with the simplest
case, $\nabla_\mu T^{\mu}_{\nu} =\delta^r_\nu k$, for a constant $k$\, \footnote{A derivation
of the type of non-conservation of energy-momentum from microscopic physics, in the spirit of~\cite{Perez:2018wlo}, is left for future work}. This is the 
choice we make in Sec.~\ref{sec:tovconstant} for constant density objects. When studying
objects described by a more realistic EoS it might also
be important to consider less simple assumptions for the non-conservation
of $T^\mu{}_\nu$. In Sec.~\ref{sec:poly} we use $\nabla_\mu T^{\mu}_{\nu} =\delta^r_\nu k \rho(r)$ for two reasons: first, because it is a
straightforward generalization of the constant violation used for constant
density objects, and second, because in a preliminary study we found that a 
constant violation does not allow for objects with compactness higher than
the compactness of GR solutions for the same density and equation of state. Thus, if UG with non-conservation of $T^{\mu}_{\nu}$ is taken
seriously, observations of highly compact objects could rule out a constant
violation of energy-momentum conservation.

%%%%%%%%%%%%%%%%%%%%%%%%%%%%%%%%%%%%%%
\section{Constant density objects in UG}\label{sec:tovconstant}
%%%%%%%%%%%%%%%%%%%%%%%%%%%%%%%%%%%%%%

We explore solutions for objects with constant density in UG. This simple scenario is always a good starting
point in the study of stellar dynamics from where some physical intuition can be drawn, even if 
we can only access the solutions numerically.

As stressed earlier, the system of equations obtained from variation of the action in UG is under-determined. In order to get a closed system, and at the same time to control the violations of conservation of $T_{\mu\nu}$, we supplement the set of equations with 
\begin{equation}
\nabla_\mu T^{\mu}{}_\nu = \delta^r_\nu k\, ,\label{eq:noncon} 
\end{equation}
for some constant $k$ that parameterizes the
non-conservation of the energy-momentum tensor of matter. Given the symmetries of our set-up, the time and angular components of $\nabla_\mu T^{\mu}{}_\nu$ vanish identically. Then, the fact that the non-conservation of energy-momentum happens only in the radial direction is not an assumption but a consequence of the symmetries of the system, what we are assuming is that this non-conservation is constant. All in all, we solve numerically
$\xi_{00}, \xi_{11}$, and \eqref{eq:noncon}
for the functions $f(r), h(r)$ and $p(r)$. 
Initial conditions are set at a small radius $r_i$ by Taylor expanding and solving
these equations near $r=0$. We find that these
initial conditions depend  on the constant density $\rho_0$, the central value
of the pressure $p(0)$, the parameter $k$, and the
second derivative of the pressure at the origin, $p''(0)$. The presence of $p''(0)$ in the initial 
conditions is a difference with respect to GR, and can be related to the additional integration
constant of UG. In particular, one can check that the initial conditions for $k=0$ only 
coincide with those of GR if $p''(0)$ takes the value dictated by the constant density solution of GR: in general, setting the additional integration constant of UG to zero means choosing $p''(0)$ in such a way that it coincides with its value in the GR solution, this is how we fix $p''(0)$. 
Another thing to note is that the first derivative of the pressure does not vanish at $r=0$, indeed it is equal to $k$, this is a consequence of the type of violation of energy-momentum conservation that we impose. More details can be found in
Appendix~\ref{app:cd}.

\begin{figure}
\includegraphics[width=0.75\textwidth]{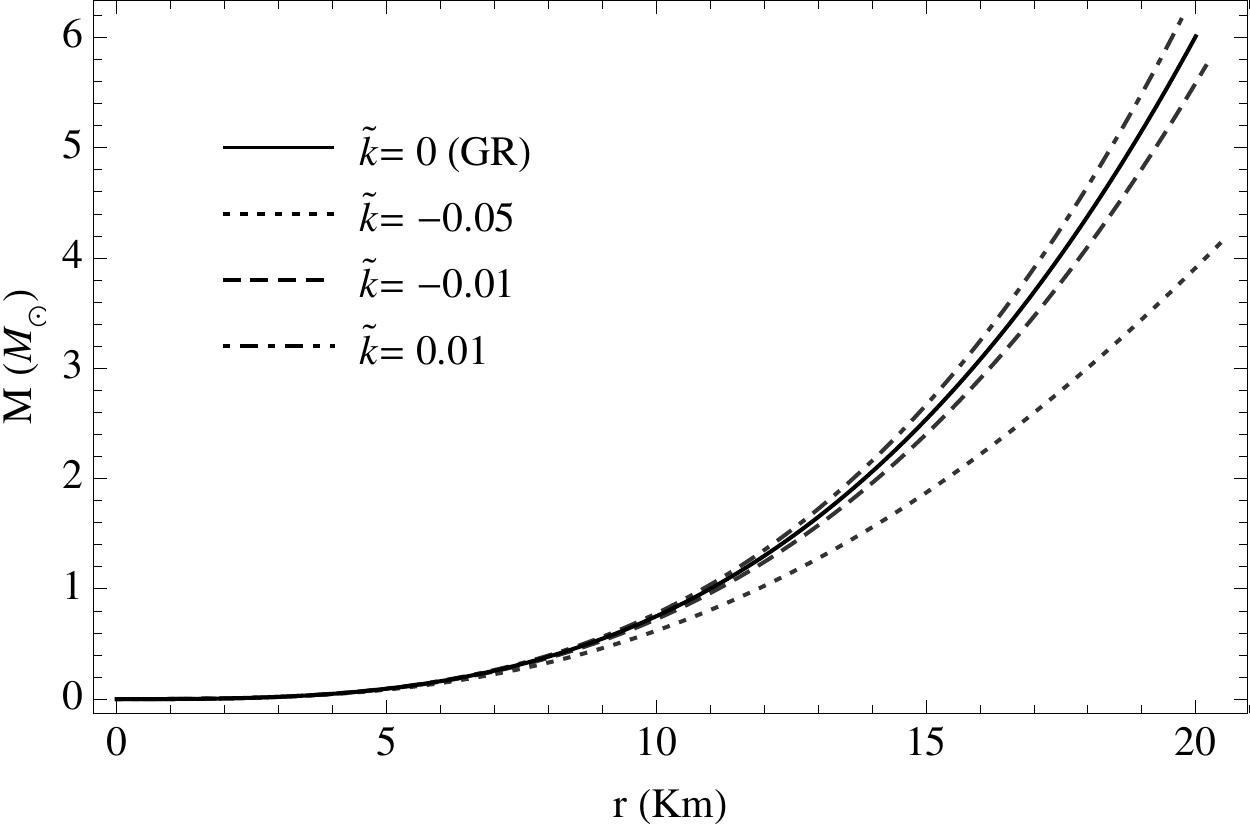}
\caption{Mass as a function of radius for configurations with
different amounts of non-conservation of $T_{\mu\nu}$. The dimensionless constant 
$\tilde k$ is defined by convenience as 
$k = c^2 (1.66\times 10^{14}\text{gr/cm}^3)(10^{-5}\text{cm}^{-1})\, \tilde k $. The numerical factors are typical density and length scales of 
astrophysical compact objects.  }
\label{fig:massesconstantrho}
\end{figure}
Fig.~\ref{fig:massesconstantrho} shows our results for the masses of constant density configurations with different values of $k$. The GR solution corresponds to $k=0$ and it is shown in solid line. Notice that sizable changes in the mass occur without large modifications in the radius of the configuration. This has interesting consequences for the compactness of these objects, defined as the dimensionless ratio $C = G\mathcal M_{UG}(R_s)/R_s$, where $\mathcal M_{UG}(R_s)$ is the mass function defined in~\eqref{eq:hmass} evaluated at the radius of the star, $R_s$. The GR solution displayed in Fig.~\ref{fig:massesconstantrho} has compactness
$C \approx 0.4435$, just below the Buchdahl limit~\cite{Buchdahl:1959zz} $C=4/9$ that comes from requiring the pressure to be finite at the center of the star. As can be inferred from the same figure, the solution with $k=0.01$ has a larger compactness, and indeed we find $C_{0.01} = 0.4635$, where we introduced the notation $C_k$ to indicate that $C$ is computed for a configuration with
a given value of $k$. 

In view of these results, it is worth exploring the 
region $k>0$ in more detail. To do so we parameterize the density in terms of the critical density in GR for uniform distributions of mass:
\begin{equation}
\rho_{0} = a \frac{8}{3 (R^{GR}_s)^2 \kappa} \equiv a \rho_{crit}\, ,
\end{equation}
for a constant $a$. The initial conditions are thus determined in terms of $a$ and $R^{GR}_s$, the latter can be fixed as the radius of the GR configuration by choosing appropriate values for $p(0)$, 
\begin{equation}
    p(0) = \frac{\rho _0 \left(-\sqrt{3}+\sqrt{3-\kappa  (R^{GR}_s{})^2 \rho _0}\right)}{\sqrt{3}-3 \sqrt{3-\kappa  (R^{GR}_s{})^2 \rho _0}}\, .
\end{equation}
For $k\neq 0$ we do not have an explicit relation between the radius of the star and the initial conditions at $r=0$, so even though we use $R^{GR}_s$
in the initial condition for the pressure, the radius of the star is $R_s \neq R_s^{GR}$. As we explained above, $p''(0)$ is chosen in such a way that for $k=0$ the GR solution is
recovered, i.e., we set to zero the additional integration constant of UG. For given $a$ and $R_s^{GR}$, the only free
parameter in the initial conditions is $k$, and by the discussion after Fig.~\eqref{fig:massesconstantrho} we are interested in $k>0$. 

\begin{figure}
\includegraphics[width=0.75\textwidth]{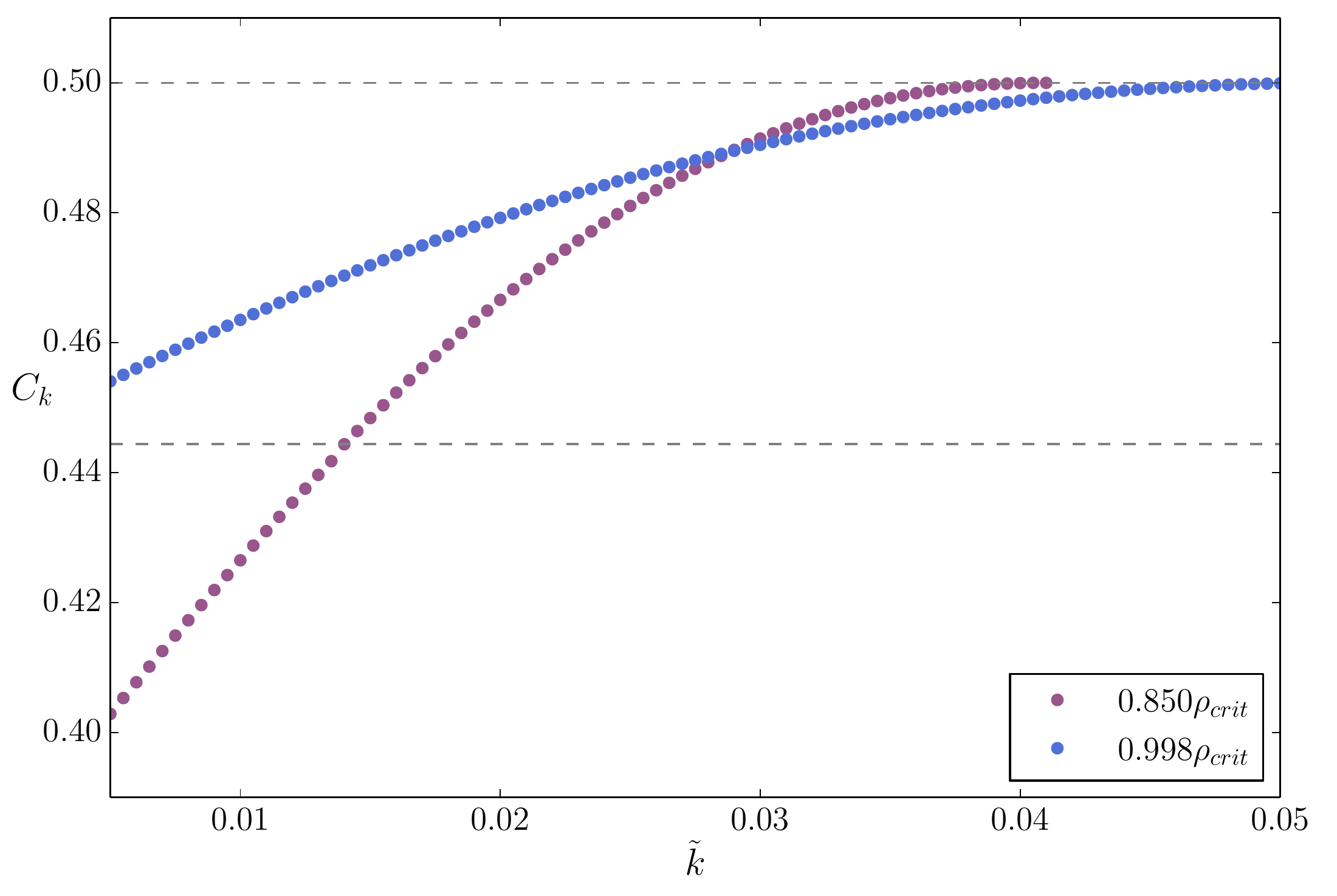}
\caption{Compactness as a function of the non-conservation of $T_{\mu\nu}$
for objects with two fixed densities near the critical density of GR solutions. The dashed line at $C=4/9$ represents the GR limit for the compactness of constant density objects, achieved only for objects with
$\rho = \rho_{crit}$. In UG, objects with $\rho<\rho_{crit}$ can surpass
this limit if the non-conservation of $T_{\mu\nu}$ is large enough.}
\label{fig:cnessrhocrit}
\end{figure}

Fig.~\ref{fig:cnessrhocrit} 
shows the compactness of configurations with $a = 0.850$
and $a = 0.998$, and values of $k$ between $0.005 \leq k \leq 0.05$. These
results confirm that solutions in unimodular gravity 
can go well beyond the Buchdahl limit $C=4/9$. Interestingly, they approach asymptotically to the compactness of a Schwarzschild black hole, $C=1/2$. For
$a=0.850$ and $k\geq 0.04$, instead of a smooth approach  $p(r) \to 0$ as $r$ approaches some value that would correspond to the radius of the compact object, we find $dp/dr \to -\infty$ at some finite radius, similar to the radius of configurations with lower $k$. It is not clear whether this is a numerical problem or a physical limit on the size of $k$. In any case, this limit is beyond 
the values of $k$ that give us $C_k \approx 0.5$, so that we do not expect it to be observationally relevant, in the sense that any observed compactness higher than $4/9$ would be
sufficiently interesting already.

In the next section we explore how the properties of constant density configurations change when we consider a still simple but more realistic approximation to the equation of state  of compact stars.

%%%%%%%%%%%%%%%%%%%%%%%%%%%%%%%%%%%%%%
\section{Polytropic stars in UG}\label{sec:poly}
%%%%%%%%%%%%%%%%%%%%%%%%%%%%%%%%%%%%%%
In this section we build upon the previous results in order to study compact objects described
by a polytropic equation of state,
chosen in such a way that GR
configurations with masses and radii in
the range of neutron stars are obtained. 
As before, we use 
a diagonal energy-momentum tensor $T^{0}_{0} = -\rho(r), T^i{}_{i} = p(r)$, with 
\begin{align}
    \rho(r) & = \rho_0 \left(\chi(r) + \frac{K}{\Gamma - 1} \chi(r)^\Gamma \right)\, , \\
    p(r) & = K \rho_0 \chi(r)^\Gamma\, ,
\end{align}
where $\chi(r)$ is a dimensionless function, while $K$
and $\gamma$ are the free parameters of the EoS.
These parameters are determined by requiring that the properties of the resulting configuration in GR match a realistic equation of state, in particular, $K= 0.0225$ and $\Gamma = 2.34$ are compatible
with the masses of PSR J1614-2230 ($1.97 \pm 0.04M_{\odot}$ \cite{Demorest:2010bx}) and PSR J0348+0432
($2.01 \pm 0.04M_\odot$ \cite{Antoniadis:2013pzd}), two of the most massive
neutron stars (NS) observationally confirmed to date\footnote{Recently, the mass of PSR J2215+5135 has been estimated to be around $2.27 \pm 0.17M_\odot$\cite{Linares:2018ppq}, but this result depends on the orbital inclination, which has not been independently confirmed.}.

In addition to the EoS, in order to close the system of equations we decide to assume 
a type of violation of conservation of $T_{\mu\nu}$. Following the discussion in Sec.~\ref{sec:vio},
%\textcolor{red}{Appendix~\ref{app:vio}} we comment
%on different choices for this violation, but 
here
we focus on 
\begin{equation}
\nabla_\mu T^{\mu}{}_{\nu} = \delta^r_\nu k \rho(r)\, .
\end{equation}
This is a generalization of the constant violation assumed in the case of constant density. 

We look for solutions numerically, setting initial
conditions at a small radius $r_i$ by solving
the equations of motion in a Taylor expansion 
around $r=0$. In contrast to GR, where
the initial conditions depend only on
$\chi(0)$, here they depend on
$\chi(0)$, $\chi''(0)$ and $k$. The value of 
$\chi''(0)$ is related to the additional integration constant of UG -- the one associated with a cosmological constant; to set this contribution equal to zero we fix $\chi''(0)$
to be the same as in the GR solution. Thus, 
the only free parameters of our solution are 
$\chi(0)$ and $k$. 

\begin{figure}%frompolytropes
\includegraphics[width=0.48\textwidth]{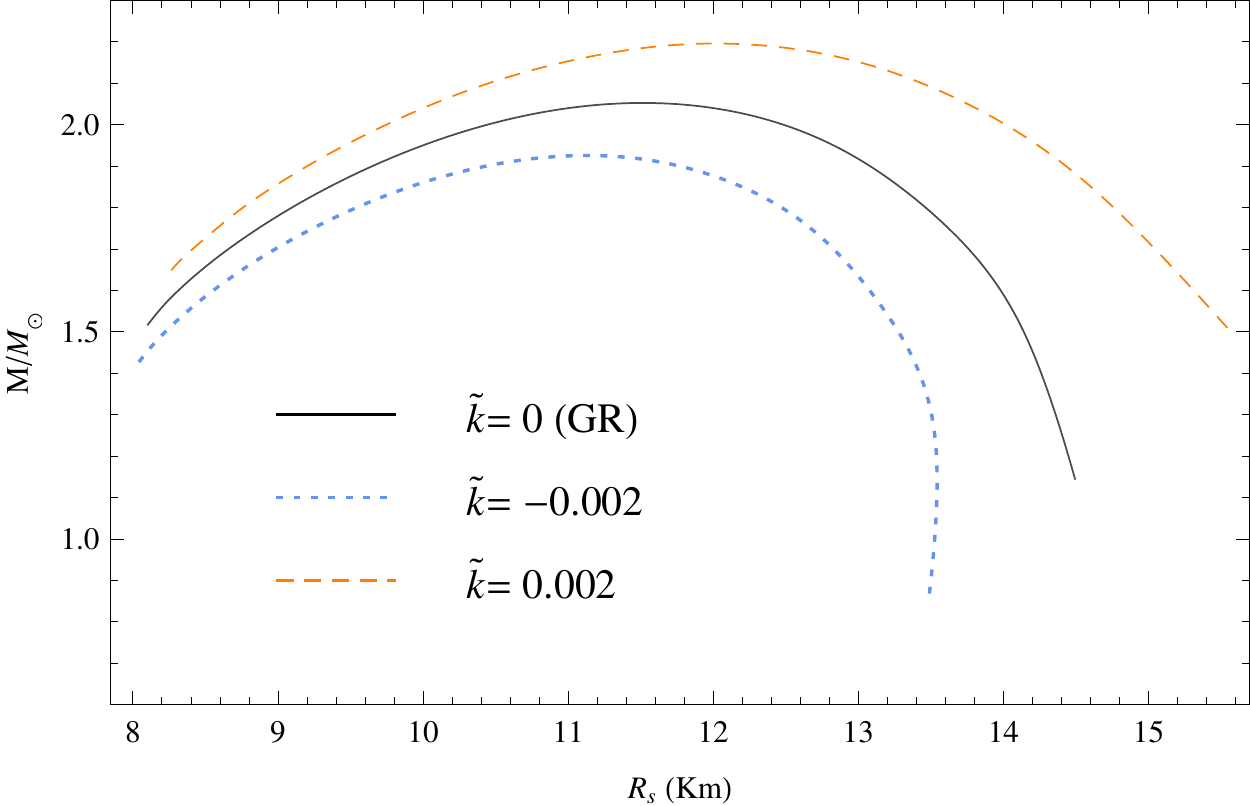} \ 
\includegraphics[width=0.48\textwidth]{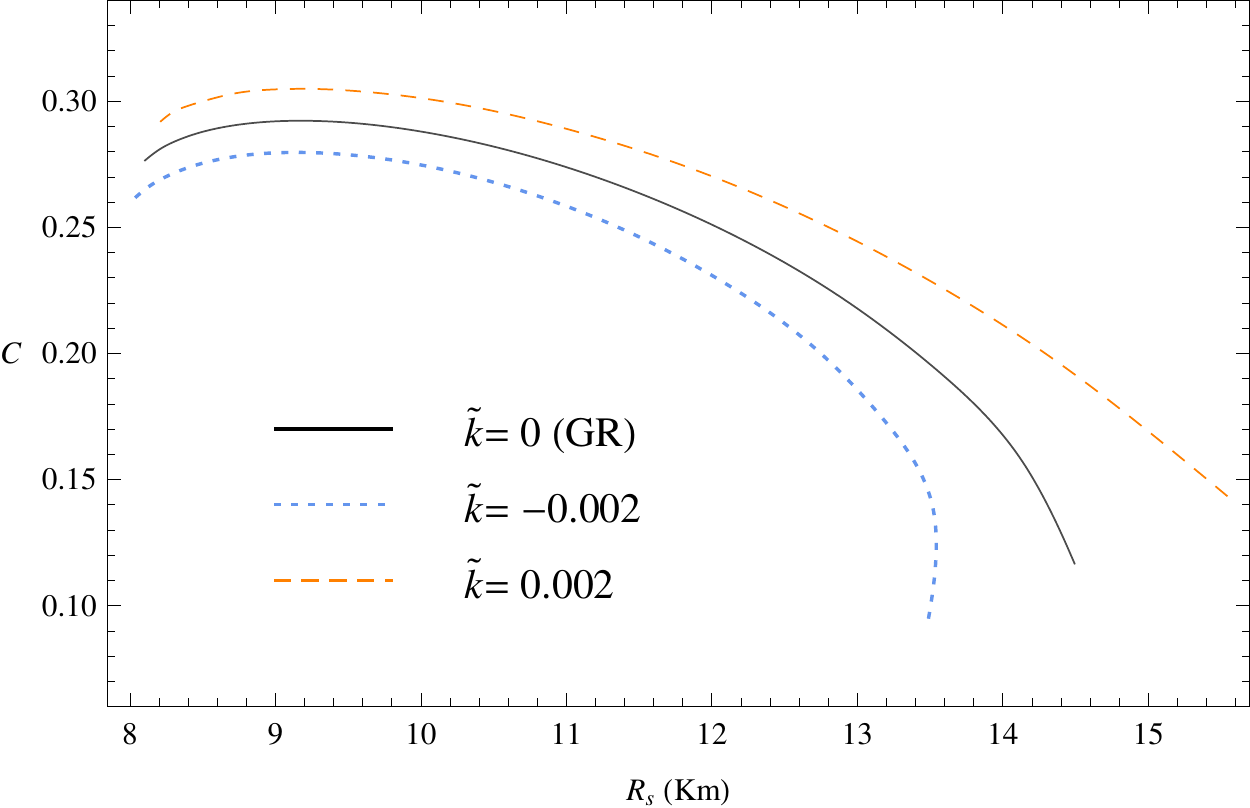} 
\caption{Mass-radius curves and compactness for compact polytropic
stars, with non-conservation of energy-momentum parameterized by the dimensionless constant
$\tilde k = (1.2\times 10^5\text{cm}) k$. Notice that the 
changes in the properties of the star due to $\tilde k$ get smaller
as the matter density increases. }
\label{fig:mrandcnesspoly}
\end{figure}

Fig.~\ref{fig:mrandcnesspoly} shows our results for polytropic configurations with $k=0$ -- which recovers GR, $k=-0.002$ 
and $k=0.002$. 
The left panel shows the mass-radius curves for equilibrium configurations with central densities in the range $10^{14}-10^{19}\textrm{\, kg/m}^3$. Assuming the same values
of $K$ and $\Gamma$ for every $k$, we find that the
mass-radius curves for negative (positive) $k$ lie below (above) the GR curve. Similar results are found for the 
compactness of these configurations, shown in the right panel
of Fig.~\ref{fig:mrandcnesspoly}. These results show that the GR solution is continuously recovered as $k\to 0$, but relatively large deviations in the compactness of low density objects appear even for small breaking of conservation of $T_{\mu\nu}$; therefore, this conservation breaking could be constrained, e.g., by
observations of the compactness and tidal deformabilities of NS (see, e.g. \cite{Most:2018hfd}). Furthermore, it is important to highlight that we generically find solutions with masses higher than the
ones allowed in GR  by the static equilibrium criterion
$dM/d\rho_c>0$ \cite{Haensel:2007yy}: stars with such masses, if observed, could hint towards modifications of the theory of gravity. Similar effects are found in scalar-tensor and vector-tensor gravity \cite{Cisterna:2015yla,Silva:2016smx,Maselli:2016gxk,Chagoya:2017fyl,Chagoya:2018lmv}.
Also in connection to other modified gravity models,
notice that the deviations due to UG shown in Fig.~\ref{fig:mrandcnesspoly} get smaller as the
central density of the stars increases, this behaviour
is reminiscent of screening mechanism (see, e.g.~\cite{Chagoya:2014}), we speculate that it can be a consequence of the metric non-linear relation and the
non-conservation of $T_{\mu\nu}$ implied by eq.~\eqref{eq:enmom}.

\begin{figure}%frompolytropes.nb
\includegraphics[width=0.48\textwidth,height=5.5cm]{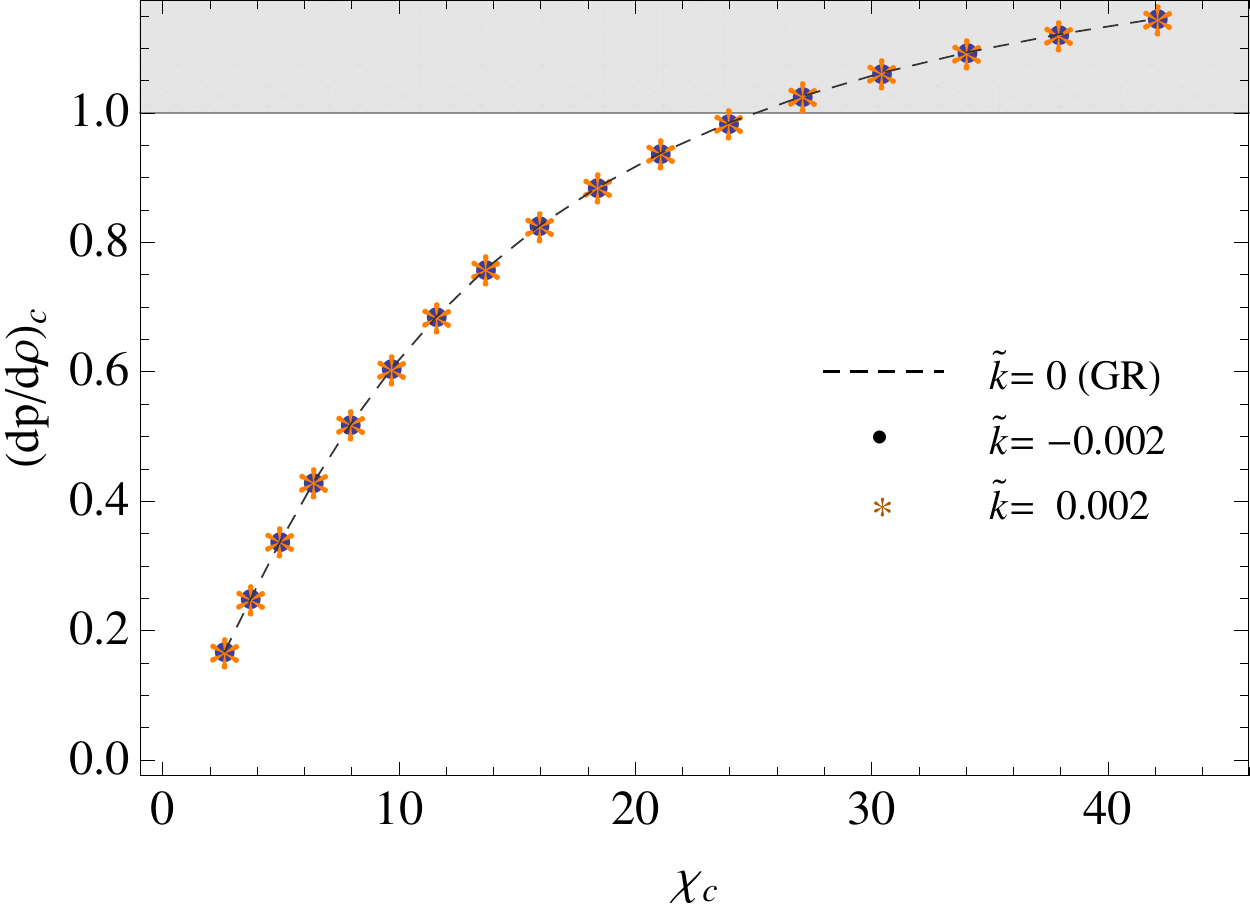} \
\includegraphics[width=0.48\textwidth,height=5.5cm]{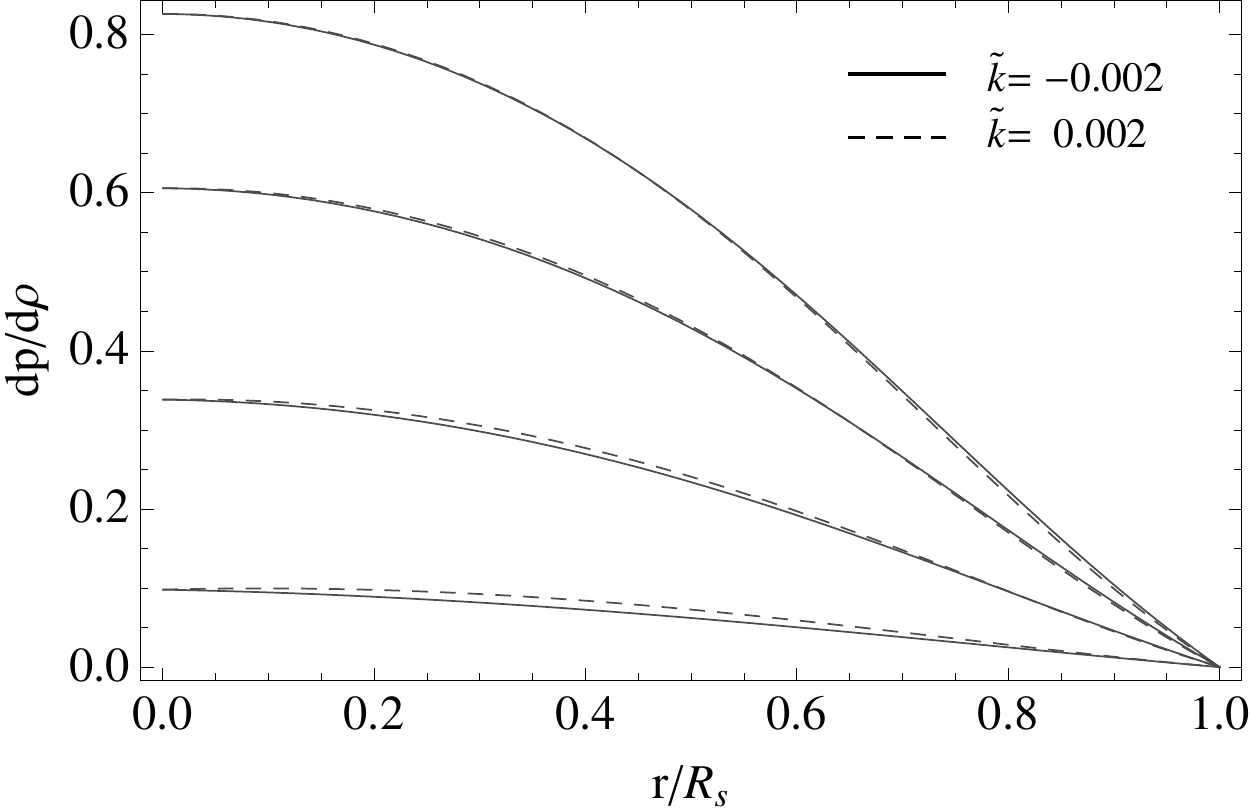}
\caption{Left panel: sound speed at $r=0$ for polytropic stars with different amounts of non-conservation of $T_{\mu\nu}$. In all cases,
the results agree with GR. Right panel: sound speed inside the star for
four different central densities. The GR curves (not displayed)
 lie between the curves for $\tilde k = -0.002$ and $\tilde k=0.002$. Small deviations from GR happen inside the star.}
\label{fig:soundspeed}
\end{figure}

Other constraints could be imposed by studying the sound speed 
$c_s^2 = dp/d\rho$ of our solutions. Causality requires $dp/d\rho\leq 1$, and also we should have $dp/d\rho\geq 0$. However; we find that the maximum $c_s^2$ in every solution
is attained at the center of the star and it depends very weakly on the value of $k$. This is shown in the left panel of Fig.~\ref{fig:soundspeed}, where the shaded area is the region excluded by causality. The solutions in that area are already
excluded by the static equilibrium criterion, so that no new constraints arise from the sound speed. The right panel
of Fig.~\ref{fig:soundspeed} shows that the changes to the
sound speed inside the star induced by $k$ are also small.

By construction, for $k=0$ the maximum mass of the configurations
shown in Fig.~\ref{fig:mrandcnesspoly} is about the observed
$2 M_\odot$ limit. For negative $k$, the maximum mass is reduced.
Although the parameters $K$ and $\Gamma$ of the EoS can be readjusted 
in such a way that the observed maximum mass limit is recovered, 
these changes to the EoS also affect other quantities. 
For instance, let $\mathcal M$ be the observed maximum mass
 and $\mathcal R$ its radius computed in GR with a polytropic equation of state. In UG with $\tilde k = -0.01$ we can obtain a mass-radius curve
 with maximum at $(\mathcal R, \mathcal M)$ (i.e. we require that the 
 radius of the maximum mass configuration is the same in GR and in UG)
 by setting the polytropic parameters $K\approx 0.0175$ and $\Gamma\approx 2.88$. However, when
 we study the sound speed of these objects we find that the limit
 $c_s^2 \leq 1$ is violated before reaching the maximum mass. Thus, we learn
 that by combining conditions on the mass, radius, and causality of the 
 solutions we can constrain the size of $\tilde k$ even if some freedom
 is allowed in the EoS: neutron stars in UG, compatible with 
 $\mathcal M\approx 2M_\odot$ and $\mathcal R \approx 12$~Km
 and described by a polytropic equation
 of state, require $\tilde k > -0.01$. Tighter constraints are not yet possible
 due to the observational uncertainty of the mass-radius curves (see, e.g.~\cite{Ozel:2016oaf}).
 For more realistic EoS, whose parameters are 
 determined from effective theories and assumptions on the behaviour of nuclear matter, there are two possibilities. First, the parameters of the EoS get modified once the microscopic non-conservation of energy-momentum
 is taken into account, in this case the situation would be similar to our discussion above for the polytropic equation of state. Second, the parameters of the EoS remain unchanged. An interesting consequence of this
 scenario 
 would be that some equations of state previously discarded for not reproducing $\mathcal M$ could be revived by assuming a minimum, positive 
 value for $\tilde k$. This could be the case, for instance, for the 
 BSk19 EoS~\cite{PhysRevC.82.035804} (see also~\cite{Potekhin:2013qqa}), which predicts a maximum mass of $1.86M_\odot$. Comparing with
 Fig.~\ref{fig:mrandcnesspoly} we speculate that $\tilde k$ of order $10^{-3}$
 can lift this value to approximately $2M_\odot$, thus making this EoS viable. In order to analyze these possibilities, a detailed study 
 of the different equations of state proposed in the literature and their
 underlying assumptions is required. Such a study is beyond the scope of this work.

Another quantity of interest is the surface redshift, i.e., the 
gravitational redshift of emission lines originating near the surface
of the star. As stated in eq.~\eqref{eq:surfz}, this redshift is closely related to the compactness. For the objects shown in Fig.~\ref{fig:mrandcnesspoly} we find redshifts modified by about $10\%$
with respect to their GR value. For instance, a polytropic object with a radius of $12$Km has $z\approx0.42$ in GR, $z\approx 0.36$ in UG with 
$\tilde\kappa = -0.002$ and $z\approx 0.47$ in UG with $\tilde\kappa = 0.002$. These changes are within the lower and upper bounds
on the redshifts expected in GR~\cite{1984ApJ...278..364L}, and therefore
do not offer an observational test of UG with $-0.002\lesssim z\lesssim 0.002$.

To conclude our discussion of the structure of the star, 
let us comment on the properties of the pressure at $r=0$.
In contrast to GR, where the equilibrium equations demand
$p'(0) = 0$, the solutions displayed in this section
have $p'(0) \neq 0$ -- indeed $p'(0) = k \rho(0)$, thus, for positive $k$ the maximum pressure is not necessarily the
pressure at $r=0$.  Also, the second derivative of 
the pressure at $r=0$ is not necessarily negative as is the case in GR but can become positive if $k$ is sufficiently large, this can be seen by exploring the perturbative solutions
near $r=0$. Nevertheless, for the stable solutions shown here
we always have $p''(0)< 0$, indicating that the profiles of $p(r)$ near $r=0$ are concave downwards, and even if $k>0$, $p'(r)$ becomes negative at a very small distance away from the origin and from there the pressure decays monotonically to zero. 

 Intuitively one would expect solutions with non-monotonic $p(r)$ to be perturbatively unstable. Thus, a valid question to ask is whether there is a
simple way to remove them from the model. Indeed this can be done for
polytropic objects by considering a non-conservation of the form
\begin{equation}
    \nabla_\mu T^{\mu}_{\nu} =\delta^r_\nu k (\rho(r) - \rho(0))\, .
\end{equation}
By expanding the equations of motion near $r=0$ it is easy to verify 
that this leads to solutions with $p'(0) = 0$. Furthermore, $p''(0)$ -- or 
equivalently $\chi''(0)$ is a free parameter that can always be chosen
negative so to guarantee that the maximum pressure is at $r=0$. 
Nevertheless, this choice is not free of problems: once again the
properties of the solutions are such that for 
configurations with $dM/d\rho>0$ the compactness is smaller than
the compactness of the corresponding GR objects. Also, the constant
term $k \rho_0$ implies that the non-conservation of $T^{\mu}_{\nu}$
is larger near the surface of the star, which is counter-intuitive in
particular for a polytropic model where $\rho(R_s) = 0$ and we expect
a smooth transition to the vacuum solution. Summing up, we found that
configurations with compactness equal or smaller than GR bounds are
generic in UG with non-conserved energy-momentum tensor, while higher
compactness is possible if we allow for the maximum density of the star
to be shifted away from the origin. It would be interesting to study
the theoretical viability of these solutions, as well as their existence
in models where the non-conservation of $T_{\mu\nu}$ is not directly 
sourced by matter but by the curvature. We leave this for future work.

%%%%%%%%%%%%%%%%%%%%%%%%%%%%%%%%%%%%%%%
\section{Gravitational Collapse in UG} \label{GC}
%%%%%%%%%%%%%%%%%%%%%%%%%%%%%%%%%%%%%%%

In this section we study the Snyder-Oppenheimer (SO) model, which is the simplest case of gravitational collapse, in the UG scenario without energy-momentum conservation, assuming a spherically symmetric collapse of dust with negligible pressure.

The metric related to this type of collapse is the well known homogeneous and isotropic line element \cite{weinberg2008gravitation} written in the form:
\begin{equation}
ds^2=-dt^2+R(t)^2\left[\frac{dr^2}{1-kr^2}+r^2d\Omega^2\right],   \label{collapse}  
\end{equation}
where $R(t)$ is the comoving radius of the star, $k$ is the curvature of the star which always must be imposed positive and $d\Omega^2\equiv d\theta^2+\sin^2\theta d\varphi^2$ is the solid angle. Using~\eqref{eq:ugeom} with the addition of a dust energy-momentum tensor ($p=0$) we have
\begin{equation}
\ddot{R}R-\dot{R}-k=-4\pi GR^2\rho_{dust}. \label{FriedmannR}
\end{equation}
Additionally, Eq. \eqref{eq:enmom} generates
\begin{equation}
\dot{\rho}_{dust}+3\mathcal{H}\rho_{dust}=\frac{\mathcal{H}^3}{4\pi G}(1-j), \label{enmomV}
\end{equation}
where $\mathcal{H}\equiv\dot{R}/R$ and $j\equiv\dddot{R}/R\mathcal{H}^3$, the last parameter is defined in order to encode the non-conservation of the energy-momentum tensor and help us to elucidate if we are not facing with spurious solutions due to the derivatives acting on the Ricci scalar, which translate into third order derivatives of $R(t)$ (in principle this should not be a problem since~\eqref{eq:enmom} is contained in the second order equations of motion, but one has to be careful when including~\eqref{eq:enmom} in the system of equations as we do in this section). Therefore, using Eqs. \eqref{FriedmannR}, \eqref{enmomV} and integrating we obtain
\begin{equation}
\mathcal{H}^2=\left(\frac{\dot{R}}{R}\right)^2=\frac{8\pi G}{3}\rho_{dust}-\frac{k}{R^2}+\frac{2}{3}\int_{t_0}^t \mathcal{H}^3\left(j-1\right)dt^{\prime} + \Lambda, \label{Frie}
\end{equation}
where $\Lambda$ is an integration constant. Notice that for $j=1$ and $\Lambda=0$ the traditional behavior for stellar collapse is recovered, therefore the GR limit is approached as $j\to1$, so deviations in gravitational collapse can be parameterized by $j\neq1$.

\begin{figure}
\centering
\par\smallskip
{\includegraphics[width=0.75\textwidth]{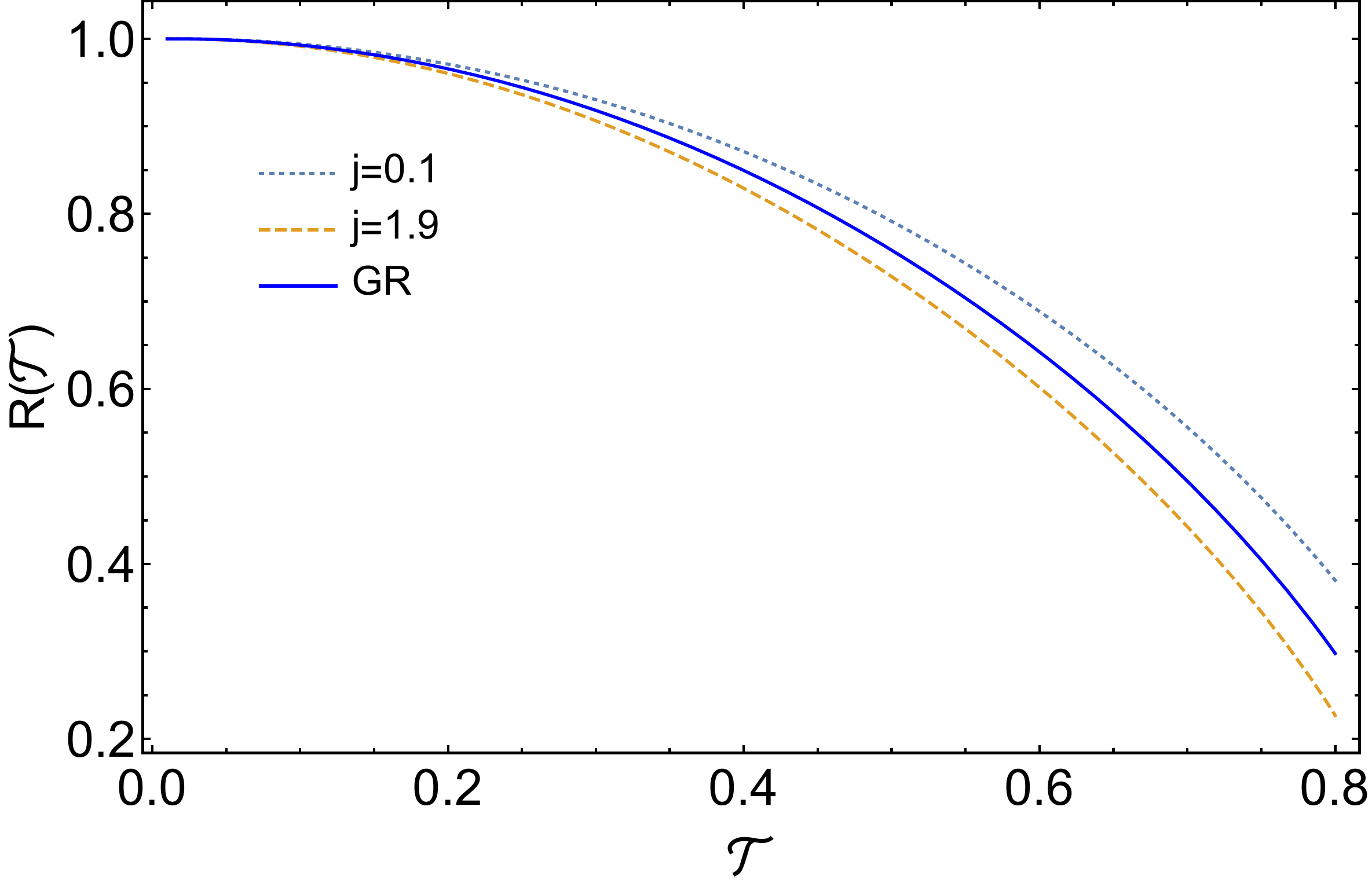}}
\caption{Numerical solutions for gravitational collapse in unimodular gravity. We choose as initial conditions: $R(0)=1$, $d{R}(0)/d\tau=0$, $\bar{\rho}(0)=1$ and $\bar{k}>0$.}
\label{Figure:Collapse}
\end{figure}

In Fig.~\ref{Figure:Collapse} we present the results of the numerical solution of Eqs.~\eqref{enmomV} and~\eqref{Frie} with  dimensionless variables $\tau\to H_0 t$, $\bar{\rho}\to4\pi GH_0^2\rho/3$ and $\bar{k}\to H_0^2 k$, where $H_0$ is an appropriate constant that has units of $s^{-1}$. In all cases we assume that the collapse initiates at a normalized radius $R(\tau)=1$ for $\tau=0$. As expected, our results show that UG differs from GR when we use different values of the parameter $j$. Here, we explore small and constant violations to energy-momentum
conservation in order to observe the differences at large values of $\tau$ in a simple model. We find that the collapse time is notably modified when we increase the presence of unimodular gravity. More exotic forms of $j$ could even stop the collapse of the star, therefore it would be interesting to study in detail how this modifies black hole formation and population.
Assuming that this process has to be very similar in UG and in GR,  we should expect $j\approx 1$, allowing the collapse of the star and only producing small violations to the energy-momentum tensor. 

It would also be interesting to study collapse in the context of structure formation. From the results
above, it would be possible for subtle differences in the presence of violations to the 
conservation of energy-momentum to modify collapse times. In this vein we suggest that the reionization \cite{Tilvi:2014,*Schenker:2014t} epoch could be an excellent laboratory to validate or refute some aspects of UG, in particular we propose a future study, through a comparison with the empirical star-formation rate proposed by \cite{Madau:2014bja} and with the strength of the 21-cm signal through the so-called differential brightness temperature $T_{21}$, in star formation eras \cite{Zaldarriaga:2003du}.

In addition, we notice that the solution of Eq. \eqref{enmomV} for values of $j\to1$ (as we expect) can be approximately written as $\rho_{dust}\approx\rho(0)_{dust}R^{-3}+{\rm Corr}$, where the corrections comes from the violations to the energy-momentum tensor, being $\rho(0)$ the central stellar density. This corrections contribute to the effective mass of the star, allowing the possibility of a larger population of NS or black holes in the Universe. Moreover, the final fate of the star strongly depends on the violations to non-conservation in UG, and also on the Chandrasekhar and Oppenheimer-Volkoff limits associated to the UG mass. Therefore, the study of the SO collapse is not enough to give a verdict on the destiny of the compact object.

%%%%%%%%%%%%%%%%%%%%%%%%%%%%%%%%%%%%%%
\section{Discussion and Conclusions} \label{Sec:CO}
%%%%%%%%%%%%%%%%%%%%%%%%%%%%%%%%%%%%%%
This paper presents a systematic study of static, spherically symmetric solutions in unimodular
gravity with non-conserved energy-momentum tensor. This non-conservation has relevant
consequences for the cosmological constant problem. However, to our knowledge, it has not been 
exhaustively studied in the strong gravity regime. Here we report some progress in this direction: 
we address issues regarding the choice of coordinate system, showing that a coordinate transformation from
unimodular to standard spherically symmetric coordinates is possible in presence of matter, we study compact and polytropic configurations of matter
in detail, and we analyze a simple model for gravitational collapse. In addition,
in Appendix~\ref{app:rn} we revisit the Reissner-N\"ordstrom solution, which had been
claimed to incorporate effects not accounted for in GR, even under the assumption that the
energy-momentum tensor is conserved~\cite{Chaturvedi:2016fea}. This would contradict the 
classical equivalence between GR and UG. We clarify this by pointing out that the solution reported
in~\cite{Chaturvedi:2016fea} makes use of results that only hold in a different version of UG, 
known as density-metric unimodular gravity~\cite{Abbassi:2007bq}, and we rederive the RN solution, finding full compatibility with GR.

Let us summarize our main results. We start by finding the TOV equation in UG under a 
specific ansatz that allows for an analytic treatment of the field equations.
We show a comparison between the gravitational energy $V/M$ in GR and in UG, obtaining that the ansatz $R_{tt}=0$ leads to positive values that indicate non-physical results in the case of constant energy density. Furthermore, different stellar compactness are obtained due to the presence of 
a constant related to the ansatz chosen to close the UG system of equations.

For constant density objects we parameterize continuous deviations from GR by choosing
an appropriate ansatz for the type of non-conservation of $T_{\mu\nu}$. We find that their 
compactness goes well beyond the Buchdahl limit as this non-conservation increases, 
and approaches asymptotically to the compactness of a black hole.

We have also studied neutron stars described by a polytropic EoS. We find that
the type of violation of $T_{\mu\nu}$ becomes relevant. We focus on a choice that allows
for objects with higher compactness than their GR counterparts, but this comes with the 
peculiarity that the maximum pressure of the star is shifted away from the origin by a
distance related to the size of the violation to energy-momentum conservation. This is in stark contrast with GR, where a monotonically
decreasing pressure is guaranteed by the equilibrium TOV equations. Similarly, in a generic class of modified gravity models the equations do admit solutions where the pressure increases with $r$ near the origin, but it was shown that a complete solution does not exist, i.e., that the
pressure never turns from increasing to decreasing as a function of $r$~\cite{Saito:2015fza}. The existence of this type of solutions in UG is thus 
a novel prediction. A detailed 
study of the stability of these solutions, left for future work, is a promising tool
to constrain this model. Furthermore, by combining causality and conditions on the mass
and radius of the maximum mass configuration we are able to set a bound on 
the parameter that controls the non-conservation of energy-momentum.

Finally, we revisit homogeneous and isotropic gravitational collapse in UG. The non-conservation of energy-momentum is encoded in the fluid equation through the $j$ parameter. We show that even if we take this
parameter close to its GR value, the solutions exhibit modifications to the collapse time that
could impact the process of black hole formation, providing another way to constrain the violations
to energy-momentum conservation. We discuss that the growth of structure in our Universe could be also affected, specifically in the reionization epoch \cite{Tilvi:2014,*Schenker:2014t}, modifying the population of collapsed objects (white dwarfs, NS or black holes). In this vein, we propose that in future works, the results must be extended and compared with the empirical star-formation rate, resulting in important consequences for the observed peak of cosmic star formation history at $z\approx2$ \cite{Madau:2014bja}; not less important could be a profound study of 21-cm signal through the $T_{21}$ temperature which is not only sensitive to the Hubble expansion but also to the temperatures that are intimately related with the stellar formation and population \cite{Zaldarriaga:2003du}.

In summary, the results presented in this work suggest several scenarios where the 
non-conservation of energy-momentum allowed in UG could be constrained. 
We have chosen to analyze simple forms of non-conservation, showing effects that 
we expect to be generic for other choices, like the modifications to the compactness of
neutron stars and the change in gravitational collapse times. In addition, we remark that UG theory can be considered as a particular case of $f(R,T)$ theories as reported in \cite{Haghani:2017vqx}. In this study, the authors adopted a Lagrangian, which contains a linear combination of Ricci scalar and a trace of the energy-momentum tensor, developing a thermodynamic and cosmological analysis, where at the end, it is possible to demonstrate that the results degenerate to the UG theory, stressing a subtle relation between $f(R,T)$ and unimodular gravity models. Finally, it would be interesting
to derive a form of non-conservation motivated by the possible discretization of space-time, as some authors have done in the study of the cosmological constant problem \cite{Perez:2018wlo,Perez:2017krv}.
This is ongoing research that will be presented elsewhere.  

\begin{acknowledgements}
%%%%%%%%%%%%%%%%%%%%%%%%%%%%%%%%%%%%%%%%%%%%%%%
We thank the anonymous referee for thoughtful remarks and suggestions. J.C.F.-U acknowledges support from IPN and UPIIZ-IPN, M.A.G.-A. and J. C. acknowledges support from SNI-M\'exico and CONACyT research fellow, M.A.G.-A. also acknowledge support from COZCyT and Instituto Avanzado de Cosmolog\'ia (IAC) collaborations. J. A. A-M acknowledge Ph.D. CONACYT fellowship.
\end{acknowledgements}

\begin{appendix}

%%%%%%%%%%%%%%%%%%%%%%%%%%%%%%%%%%%
\section{Equivalence between metrics} \label{app:eq}
%%%%%%%%%%%%%%%%%%%%%%%%%%%%%%%%%%%
A point that tends to cause confusion in UG is whether it is necessary to use a metric that fulfills the unimodular condition or a metric in traditional spherical coordinates -- or any coordinates -- can be used. The requirement that the metric determinant equals a constant is a coordinate-dependent statement, and one should prefer a statement about coordinate-independent objects, like the volume form. We also remark that the goal of the 
unimodular condition is to restrict the variations of the metric and not the metric per se. 
Nevertheless, it is good to show explicitly the equivalence between metrics in unimodular and
 in other system of coordinates. Here we do so for a FLRW metric. 

Let us compare physical results derived for a FLRW line element written in the form $ds^2=-dt^2+a(t)^2d\vec{x}^2$ and for a metric that fulfills the unimodular condition, used in Alvarez et al.~\cite{Alvarez:2015sba} to study a cosmological scenario in UG. This 
metric reads
\begin{equation}
    ds^2=-b(\tau)^{-2/3}d\tau^2+b(\tau)^{1/2}d\vec{x}^2. \label{ucondition}
\end{equation}
Notice that Eq.~\eqref{ucondition} can be constructed from the standard FLRW line element 
via a change of variables $a\to b^{1/4}$ and $dt\to b^{-3/4}d\tau$.

Assuming a perfect fluid energy-momentum tensor and using the field equations of UG (see Eq.~\eqref{eq:ugeom}) we have
\begin{equation}
\frac{b^{\prime\prime}}{b}-\frac{1}{4}\left(\frac{b^{\prime}}{b}\right)^2=-16\pi Gb^{-3/2}(\rho+p),   
\end{equation}
where primes indicate derivatives with respect to $\tau$. This is the Friedmann equation under the unimodular condition, but the physical interpretation in this form is not straightforward. However, if we return to the $a(t)$ and $t$ variables to recover the traditional FLRW line element, we have: $\dot{H}=-4\pi G(\rho+p)$,
which is the same equation used previously by several authors \cite{Ellis,Gao:2014nia}. Therefore, the result is independent of using the FLRW metric or the metric of Eq.~\eqref{ucondition}. The essential point is, which metric provides the best insight into the physical interpretation of the results.

%%%%%%%%%%%%%%%%%%%%%%%%%%%%%%%%%%%
\section{Ansatz election} \label{Cdem}
%%%%%%%%%%%%%%%%%%%%%%%%%%%%%%%%%%%
\begin{lm}
Considering the metric \eqref{eq:sssmetric}, there is a chart such that $R_{tt}=0$ and $R_{\theta\theta}\neq 0$.
\end{lm}
\begin{de}
Let $p$ a point in the space-time with metric \eqref{eq:sssmetric}, then there is a chart $\mathcal{U}^\prime$ containing $p$, where the first partial derivative of the metric tensor vanishes at the point. Also, since $h(r)=1-2M_{UG}G/\hat{r}$ and $ \partial_{\hat{r}} h\rvert_p=-2G(\frac{M^{\prime}_{UG}\hat{r}-M_{UG}}{\hat{r}^2})\Big\rvert_p=0$, we have
\begin{align}
\partial^2_{\hat{r}} h&=\frac{-2G}{\hat{r}^4}\Big[(M^{\prime\prime}_{UG}\hat{r}+M^{\prime}_{UG}-M^{\prime}_{UG})\hat{r}^2-2(M^{\prime}_{UG}\hat{r}-M_{UG})\hat{r}\Big]\nonumber\\
&=-2G   \Big[    \frac{ M^{\prime\prime}_{UG} }{ \hat{r} } -2\frac{  (M^{\prime}_{UG}\hat{r}-M_{UG} ) }{ \hat{r}^3 }\Big]\nonumber\\
&=\frac{-2G}{\hat{r}} \Big(M^{\prime\prime}_{UG}-2\partial_{\hat{r}} h \Big)
 \end{align}
and at the point $p$  
\begin{equation}
M^{\prime}_{UG}=\frac{M_{UG}}{\hat{r}}, \qquad M^{\prime\prime}_{UG}\Big \rvert_p=-\frac{1}{2G}\partial_{\hat{r}} h\Big\rvert_p =0,
\end{equation}
resulting
\begin{equation}
\partial^2_{\hat{r}} h\Big\rvert_p=\frac{4G}{\hat{r}}\partial_{\hat{r}} h\Big\rvert_p =0.
\end{equation}
 Finally, we can infer
 \begin{equation}
 R_{tt}(p)=0,\qquad R_{\theta\theta}(p)=-1+h(r)\neq 0,
 \end{equation}
which is the ansatz that we used.
\end{de}

%%%%%%%%%%%%%%%%%%%%%%%%%%%%%%%%
\section{Equations of motion for constant density objects in UG}\label{app:cd}
%%%%%%%%%%%%%%%%%%%%%%%%%%%%%%%%
In section~\ref{sec:tovconstant} we obtained numerical solutions
for constant density objects. Here we elaborate further on the
analytic treatment of the system of equations. The angular part of
the Einstein equations satisfies $\xi^\theta{}_\theta = \xi^\varphi{}_\varphi$ because of our spherically symmetric set-up,
and $\xi^t{}_t + \xi^r{}_r+  \xi^t{}_t + \xi^{\theta}{}_\theta = 0$
because of the trace-free condition.  Thus, there are only two independent equations, say
$\xi^t{}_t$ and $\xi^r{}_r$. Notice that the trace-free condition plays the
role usually played by the Bianchi Identities and the conservation of 
$T_{\mu\nu}$ in eliminating one of the metric equations of motion. Explicitly, under
ansatz~\eqref{eq:sssmetric} the independent equations read
\begin{align}
    f^2 \left(4 r h'+4 h+6 \kappa  p r^2+6 \kappa  \rho  r^2-4\right)+h r^2 f'^2-f r \left(r f' h'+2 h \left(r f''+2 f'\right)\right) & = 0\, , \\
    -2 f^2 \left(2 r h'-2 h+\kappa  p r^2+\kappa  \rho  r^2+2\right)+h r^2 \left(f'\right)^2-f r \left(r f' h'+h \left(2 r f''-4 f'\right)\right) & = 0\, ,
\end{align}
where $f, h$ and $p$ are functions of $r$ and $\rho = \rho_0$ is constant. 
In order to close the system of equations we choose to assume a type of
non-conservation of $T_{\mu\nu}$: $\nabla_\mu T^\mu{}_\nu = k_\nu$ for some
constant vector $k_\nu$ with units of density over distance. Only the radial
component of these equations is not trivial
\begin{equation}
    p f'+\rho_0  f'+2 f p' = 2 f k\, ,
\end{equation}
where $k_\nu = \delta^r_\nu k$. After a few manipulations we obtain that the
metric satisfies
\begin{align}
    \nu'(r) & = \frac{h'+\kappa  p r+\kappa  \rho  r}{h}\, , \\
    h(r) & = 1 + c_1 r^2-\frac{1}{4} \kappa  k r^3\, ,
\end{align}
with $f(r) = e^{\nu(r)}$. The GR form of the radial component is recovered when
$k=0$ and $c_1 = - \kappa \rho_0/3$. The arbitrariness of $c_1$ comes as a consequence of the additional integration constant contained in UG. The pressure
is determined by the equation
\begin{align}
  &  \kappa  r \left[8 c_1 r \left(\rho _0-k r\right)+2 \kappa  k^2 r^3-k \left(3 \kappa  \rho _0 r^2+8\right)+4 \kappa  \rho _0^2 r\right]\nonumber \\
    & +\kappa  r^2  \left(8 c_1+8 \kappa  \rho _0-3 \kappa  k r\right) p(r) +4 \kappa ^2 r^2 p(r)^2 +\kappa  r \left(8 c_1 r^2-2 \kappa  k r^3+8\right) p'(r) = 0. \label{eq:pcd}
\end{align}
As in GR, this is a Ricatti equation, but with more complicated coefficients. Writing $c_1 = - \kappa \rho_0/3 - \gamma$ we can verify that the GR solution for
a constant density object in presence of cosmological constant $\Lambda = 3 \gamma$
is recovered when $k=0$. It is interesting to note that this effective
cosmological constant is not related to the non-conservation of $T_{\mu\nu}$. 
For arbitrary $k$ we could not find exact solutions. As a complement to the
numerical analysis performed in Sec.~\ref{sec:tovconstant}, let us study the
near-origin solutions. Expanding Eq.~\eqref{eq:pcd} near $r=0$ and setting
$c_1 = - \kappa \rho_0/3$ we find
\begin{align}
    p'(0) & = k\, ,\\ 
    p''(0) & = -\frac{1}{6} \kappa (\rho_0 + p(0))(\rho_0 + 3 p(0))\, ,\\
    p'''(0) & = -\frac{1}{12} k \kappa  (7 \rho_0+15 p(0))\, .
\end{align}
Odd orders are turned on by $k$, and even orders in general do receive
modifications due to $k$ ($p''(0)$ is independent of $k$ by construction). 
If we choose a different value of $c_1$, modifications due to the extra integration constant in UG appear at
every order except $p'(0)$. As a consequence, the maximum pressure is shifted away from the origin if $k>0$. As shown in Sec.~\ref{sec:poly}, this feature 
persists when a polytropic EoS is considered. It would be interesting to explore its implications in more detail, for example, as another way to constrain $k$ by stability conditions or as a new effect non-degenerated with the EoS. Results in these directions will be reported elsewhere. 

\section{Reissner-Nordstr\"om solution in UG} \label{app:rn}
For completeness we briefly review and clarify some aspects of the 
Reissner-Nordstr\"om solution in UG, partially reported in~\cite{Chaturvedi:2016fea}. As discussed after Eq.~\eqref{eq:sssmetric},
a spherically symmetric space-time can be described in unimodular 
coordinates by the line element
\begin{equation}
ds^2 = -f(y)dt^2 +  \frac{dy^2}{r(y)^4 f(y)}  + \frac{r(y)^2dx^2}{1-x^2} + r(y)^2 (1-x^2) d\varphi^2\, ,\label{eq:ugcoords2}
\end{equation}
where $dr = \sqrt{h(r)/(r^4 f(r))} dy$, with $f(r)$ and $h(r)$ the
$g_{tt}$ and $g^{rr}$ components of the metric in spherical coordinates. In particular, for 
a Reissner-Nordstr\"om black hole we have
\begin{equation}
    f(r) = h(r)= 1 - \frac{2M}{r} + \frac{Q^2}{r^2}\, ,
\end{equation}
and $r = (3 y)^{1/3}$. Here we want to verify that this solution, supplemented
with a cosmological constant, is the only solution in UG  for an
electrically charged black hole. To this end, we insert~\eqref{eq:ugcoords2}
without any assumptions on the form of $f(y)$ and $r(y)$ in the 
equations of motion 
\begin{align}
    R_{\mu\nu} - \frac14 g_{\mu\nu} R & =  \frac14 \left( T_{\mu\nu} - \frac14 g_{\mu\nu} T \right) \, , \\
    \nabla_{\mu} F^\mu{}_{\nu}&  = 0\, , \label{eq:metveceqs}
\end{align}
where 
\begin{equation}
    T_{\mu\nu} = -\frac12 g_{\mu\nu} F^{\alpha\beta}F_{\alpha\beta} 
    + 2  F^{\alpha}{}_{\mu} F_{\alpha\nu}\, ,
\end{equation}
and $F_{\mu\nu} = \partial_\mu A_\nu - \partial_\nu A_\mu$.
Notice that -- as we do throughout this work -- we are using the 
GR energy-momentum tensor 
$$ 
T_{\mu\nu} = \frac{1}{\sqrt{-g}}\frac{\delta (\sqrt{-g} F^{\alpha\beta}F_{\alpha\beta})}{\delta{g^{\mu\nu}}}
$$
instead of the UG version $E_{\mu\nu} =  \frac{\delta ( F^{\alpha\beta}F_{\alpha\beta})}{\delta{g^{\mu\nu}}} $ used in~\cite{Chaturvedi:2016fea}. This is justified since, as shown in~\cite{Smolin:2009ti}, 
$$ T_{\mu\nu} - \frac14 g_{\mu\nu} T = E_{\mu\nu} - \frac14 g_{\mu\nu} E\, . 
$$
Neglecting the integration constant in $r(y)$ since it is an arbitrary constant
in a change of coordinates, fixing the integration constant in $f(y)$ in
such a way that $f(y\to\infty) = 1$, and using gauge invariance of 
$A_\mu$ to set to zero another integration constant, we arrive to the 
general solution
\begin{align}
r(y) & = (3 y)^{1/3}\, , \\
f(y) & = 1 + \frac{Q^2}{ ( 3 y)^{2/3}} - \frac{2 M}{(3 y)^{1/3}} + b y^{2/3}\, , \\
A(y) & =  \frac{2 Q}{(3 r)^{1/3}}\, .
\end{align}
This is nothing more than the RN solution in presence of a cosmological 
constant expressed in unimodular coordinates. This differs from the results
in~\cite{Chaturvedi:2016fea}, where additional corrections to the RN
solution are reported. To our understanding, the difference arises because in~\cite{Chaturvedi:2016fea} the equations of motion are computed by using
the results of~\cite{Abbassi:2007bq} to evaluate the curvatures appearing
in~\eqref{eq:metveceqs}, these results hold for a theory that is 
closely related but different to unimodular gravity, in which the metric
is considered as a tensor density of weight $-1/2$ instead of a tensor
density of weight $0$, for this reason this theory is dubbed 
\textit{density-metric unimodular gravity}, and as pointed out in~\cite{Abbassi:2007bq}, 
the solutions to this theory differ from the solutions to GR and to
standard UG. This subtlety is not mentioned in~\cite{Chaturvedi:2016fea}, and 
we think it is important that we have clarified it here. 
%%%%%%%%%%%%%%%%%%%%%%%%%%%%%%%%%%%
\end{appendix}

%-------------------------------------------------------------------------------------------------
%------------------------------------------------------------------------------------------------
\bibliography{librero1}

\end{document}